\documentclass[a4paper,11pt,twoside]{article}
\usepackage{mathptmx}
\usepackage{amssymb}
\usepackage{amsmath}
\usepackage[dvips]{graphicx}
\usepackage[latin1]{inputenc}
\usepackage{color}
\usepackage{multirow}

\newcommand{\EPJ} {{\rm Eur. Phys. J.} }
\newcommand{\NIM} {{\rm Nucl. Instrum. Methods\/} }
\newcommand{\NP}  {{\rm Nucl. Phys.} }
\newcommand{\PL}  {{\rm Phys. Lett.} }
\newcommand{\PR}  {{\rm Phys. Rev.} }
\newcommand{\PRL} {{\rm Phys. Rev. Lett.} }
\newcommand{\RMP} {{\rm Rev. Mod. Phys.} }

\begin{document}

\setcounter{page}{1}

   \sloppy
   \thispagestyle{empty}
   \vspace*{2.9 cm}
   \begin{center}
   {\Large  Accurate mass measurements of $^{26}$Ne, $^{26-30}$Na, $^{29-33}$Mg performed with the {\sc Mistral} spectrometer}
                    \\  \vspace {3mm}
   {\large
           C.~Gaulard$^*$, G.~Audi, C.~Bachelet, D.~Lunney, M.~de~Saint~Simon, C.~Thibault and N.~Vieira
                          \footnote{\hspace{-6pt} * ~ Corresponding author :C. Gaulard \\
                         {\it E-mail address:} gaulard@csnsm.in2p3.fr,\  {\it Telephone:} +33 1 69 15 45 57,\  {\it Fax:} +33 1 69 15 52 68}
                  }\\ \vspace {2mm}
   {\scriptsize\it
                  Centre de Spectrométrie Nucléaire et de
                  Spectrométrie de Masse, CSNSM,
                  IN2P3-CNRS\&UPS, Bâtiment 108,
                  F-91405 Orsay Campus,
                  France
                  \\
   }
   \end{center}
   \rule{30pc}{.5pt}
   {\small
   {\\\bf Abstract}

   The minuteness of the nuclear binding energy requires that mass
measurements be highly precise and accurate.
   Here we report on new measurements $^{29-33}$Mg and $^{26}$Na performed
with the {\sc Mistral} mass spectrometer at {\sc Cern}'s {\sc
Isolde} facility.
   Since mass measurements are prone to systematic errors,
considerable effort has been devoted to their evaluation and
elimination in order to achieve accuracy and not only precision.
   We have therefore conducted a campaign of measurements for
calibration and error evaluation.
   As a result, we now have a satisfactory description of the
{\sc Mistral} calibration laws and error budget.
   We have applied our new understanding to previous
measurements of $^{26}$Ne, $^{26-30}$Na and $^{29,32}$Mg for which
re-evaluated values are reported.

   }
 \\\rule{30pc}{.5pt}

   \section{Introduction}    \label{sec:intro}

   The nuclear binding energy provides information of capital
importance, not only for nuclear physics, but for related domains such
as weak interaction studies and nucleosynthesis.
   Due to the small fraction of the binding energy compared to the
overall mass, measurements of this quantity must necessarily be of
high accuracy.
   Depending on the detail of the nuclear property that needs to be
elaborated, the required relative precision ranges from $10^{-5}$
(for shell effect studies) to $10^{-8}$ (for weak interaction
studies).
   Various techniques, conditioned by the exotic nuclide production
scheme, are used for measuring masses.
   A recent review paper gives details for the
above points \cite{Lun02}.
   The {\sc Mistral} spectrometer at {\sc Isolde} offers an excellent
combination of high precision and sensitivity.
   It is particularly well adapted to the measurement of short-lived
nuclides.    The principle of  {\sc Mistral} is presented and the
calibration procedure is fully developed in this paper.

With the goal to provide not only precise, but also accurate masses,
we examine here very carefully the uncertainties introduced by the
lack of stability and reproducibility of the measurements as well as
the calibration procedures. We have chosen to consider these
uncertainties one by one, especially as they act at different steps
of the analysis. Comparison of our results with measured masses that
were not used as calibrants will show that possible residual errors
are negligible.

   Several measurements, performed under several experimental conditions
in which different ionization and/or mass separation schemes were
used, are discussed in order to explain the new calibration
procedure. We then report the new measurements of $^{29-33}$Mg and
$^{26}$Na with the detailed analysis of the data and the resulting
precise and accurate masses ("{\sc Rilis}" experiment).
   Discussion of the experiment and associated physics at $N=20$ is the subject
of another publication~\cite{Lun04}. The results of previous {\sc
Mistral} measurements of Ne and Mg isotopes ("{\sc Plasma}"
experiment) \cite{Mon00,Lun01b}, and of Na isotopes ("{\sc Thermo}"
experiment) \cite{Lun01a} are reanalyzed using the new calibration
procedure and improved values are presented here, replacing the
older ones.

   \section{Description of the {\sc Mistral} spectrometer\protect}  \label{sec:mistral}

   The {\sc Mistral} spectrometer has been described in detail by de
Saint Simon \textit{et al.}, \cite{Sai95} and by Lunney \textit{et
al.}, \cite{Lun01b,Lun01a,Lun96}.
   The mass is determined via the cyclotron frequency $f$ of an ion
of charge $q$ and mass $m$, rotating in a homogeneous magnetic field
$B$:
   \begin{equation}
      \label{eq:fc}
   f= \frac{q\,B}{2 \pi\, m}\;.
   \end{equation}

\begin{figure}
   \begin{center}
   \includegraphics[clip,width=20pc,angle=-90]{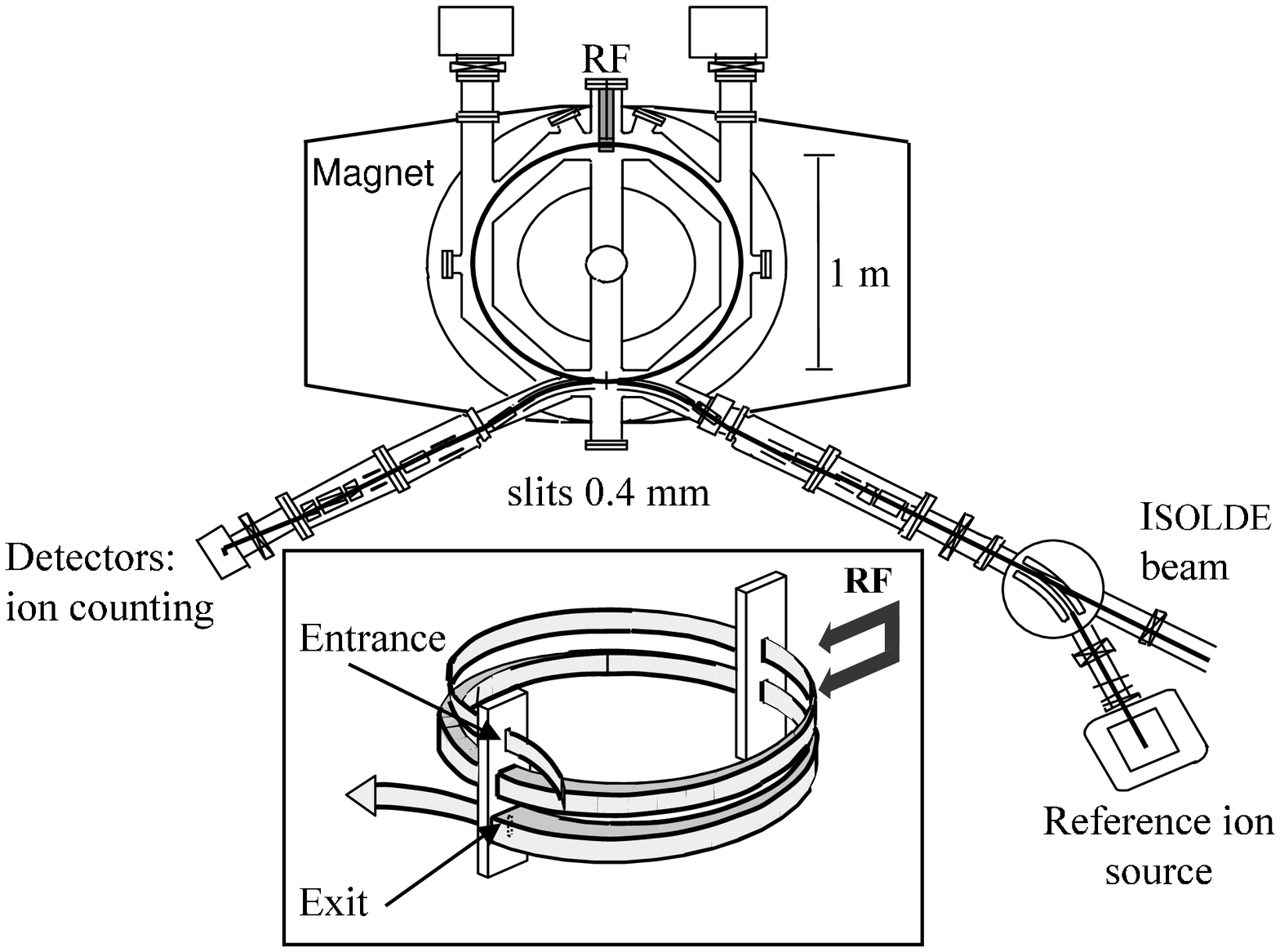}
    \caption{\small
   Layout of the {\sc Mistral} spectrometer (overhead view).
   The ion beams (coming from the right) are injected either from the
{\sc Isolde} beamline (at 60~keV) or from a reference ion source
(variable energy).
   Inset shows a view of the trajectory envelope
with the 0.4~mm entrance slit followed by the first modulator after
one-half turn, an opening to accommodate the modulated-ion
trajectories after one-turn, the second modulator after three-half
turns, and the exit slit.
   }
   \label{fig:spectro}
   \end{center}
   \end{figure}

   The layout of {\sc Mistral} is shown in Fig.\,\ref{fig:spectro}.
The ion beam is injected through the fringe field of the magnet and
focused onto the entrance slit by an electrostatic deflector.
   Then, the ions follow a two-turn
helicoidal trajectory inside the magnetic field
(Fig.\,\ref{fig:spectro}, inset) before being extracted towards an
electron multiplier for counting.
   The injection and extraction beam optics, composed of electric quadrupoles
and deflectors, are symmetric. The nominal trajectory is defined by
four 0.4~mm $\times$ 5~mm slits. At the one-half and three-half
turns inside the magnetic field, a radiofrequency modulation of the
longitudinal kinetic energy is performed. The beam is transmitted
through the exit slit of the spectrometer only if the two
modulations cancel, \textit{i.e.} when the radiofrequency $f_{RF}$
is related to the cyclotron frequency $f$ by:
   \begin{equation}
   \label{eq:frf}
   f_{RF} = (n + \frac{1}{2})\, f\;,
   \end{equation}
where the harmonic number $n$ is an integer. This harmonic number is
obtained as
   \begin{equation}
   \label{eq:n}
   n = \frac{2 \pi\, m f_{RF}}{qB} - \frac{1}{2}.
   \end{equation}
It is typically 1000-2000, therefore one needs to determine $m$ and
$B$ with an accuracy of $10^{-4}$ to calculate it. The magnetic
field $B$ is measured with this accuracy thanks to a NMR probe.
Masses, measured or estimated, are always known or predicted to
better than 1~MeV, which, above $A=10$, is always better than
$10^{-4}$.

\begin{figure}
   \begin{center}
   \includegraphics[width=18pc]{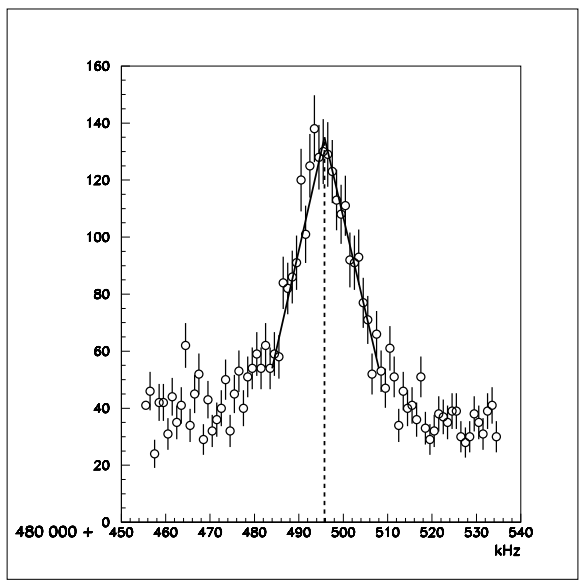}
   \caption{\small
   Accumulated transmission peak for $^{33}$Mg.
   Due to the weak production rate, a modest mass resolving power of
only about 20\,000 was used to favor transmission. For more abundant
species a resolving power of 100\,000 can be obtained.
   }
   \label{fig:33mgpic}
   \end{center}
   \end{figure}

   The ion signal recorded over a wide frequency scan exhibits
transmission peaks that are evenly spaced at the cyclotron
frequency. The frequency, $f_{RF}$, is obtained by determining the
center of symmetry of these peaks using a triangular shape fit, as
shown in Fig.\,\ref{fig:33mgpic}, \cite{Mon00,Coc88}.

   \section{Data taking and processing}  \label{sec:data}

   As mentioned above, the mass of a nuclide is directly
related to the cyclotron frequency and the magnetic field (Eq.
(\ref{eq:fc})). However, the magnetic field value cannot be
precisely measured. We thus make use of ions with a known mass
(`reference mass', $m_r$) that must follow the same trajectory as
the ion beam from {\sc Isolde}, through the same magnetic field. To
eliminate the error contributions due to long-term drifts of the
magnetic field, these beams are transmitted alternately through the
spectrometer. This requires switching rapidly (within seconds) all
the voltages. By measuring the corresponding cyclotron frequencies,
$f_r$ and $f_x$, the `measured' mass, $m_x$, can be obtained as
\begin{equation}
 \label{eq:mx}
\frac{ m_x}{q_x} = \frac{m_r}{q_r} \frac{f_r}{f_x}\;.
\end{equation}
The {\sc Isolde} and reference cyclotron frequencies, $f_x$ and
$f_r$, are determined from the frequency peak centroids according to
Eq.~(\ref{eq:frf}). The reference beam is provided by an auxiliary
ion source (Fig.\,\ref{fig:spectro}). While the magnetic field is
unchanged, the voltages of all the electrostatic elements of the
spectrometer must be set in strict inverse proportion to the
mass/charge ratio being measured:
   \begin{equation}    \label{eq:mv}
   \frac{m_{r}}{q_r} V_{r} = \frac{m_x}{q_x} V_{x}\;.
   \end{equation}

The auxiliary source can operate in surface ionization or plasma
mode, depending of the ions needed as reference. In the surface
ionization mode, alkali elements are ionized at the filament exit;
this mode is chosen when we need $^{23}$Na or $^{39,41}$K as in
references \cite{Mon00,Lun01b,Lun01a}. In plasma configuration, the
atoms or molecules to be ionized are introduced in the source by
means of an oven or a gas bottle. In an experiment to measure the
mass of $^{74}$Rb \cite{Vie02b}, $^{74}$Ge and $^{76}$Ge were used
as reference. For the latest Mg measurements (in 2001) reported here
for the first time, air was introduced in the source in order to get
the nitrogen ($^{14}$N$^{14}$N) needed as reference. In the case of
the 2002-2003 $^{11}$Li acquisition, $^{10}$B and $^{11}$B were used
as references \cite{Bac04}.

The radioactive beam is produced by {\sc Isolde} at {\sc Cern} in
nuclear reactions induced by 1 or 1.4 GeV protons from the
PS-booster. For all three experiment presented here, the target
element was uranium. Then the radioactive atoms are ionized inside a
chamber by surface ionization ({\sc Thermo}), plasma-discharge ({\sc
Plasma}), or resonant laser ionization ({\sc Rilis}). Once ionized,
the beam is accelerated to 60~keV and mass separated. Detailed
descriptions of the ISOL technique are provided in
\cite{Lun02,Kug00}.

Since the proton beam is pulsed, the time structure of the
short-lived radioactive beams implies that the radiofrequency must
be scanned one step per proton pulse. To avoid the influence of
eventual drifts, the order of the frequencies is randomly selected.
The transmission peak is then reconstructed at the end of a complete
`cycle'. This mode gives also the time dependance of the beam
intensity after the proton pulse. This feature is very useful, since
the fall-time of this so-called release curve, reflects the
half-life of the nuclide in question. It helps in identifying
whether isobaric contamination might be present and allows a time
gate to be set in order to reduce background. For the reference
mass, fast sequential scans are performed, at least one per `cycle'.
This procedure eliminates the long term fluctuations of the magnetic
field.

Finally, a single `measurement' is obtained by repeating identical
`cycles'. Typically we accumulate 10-20 cycles for one measurement
of a known nuclide. If the peak population in each cycle is not
sufficient to allow a fit (this is the case for the most exotic
nuclear species), the successive spectra are summed before being
fitted , and the interspersed reference masses measured during the
same time are averaged. The example of $^{33}$Mg, shown in Fig.
\ref{fig:33mgpic}, is a sum of 145 cycles. If the statistics are
sufficient to allow a fit, each cycle gives a mass value. These mass
values are averaged and a normalized $\chi$-value for each
measurement (Birge ratio) is derived. Figure~\ref{fig:chi4}-a shows
the $\chi$-values for each of the 23 mass measurements performed
during the {\sc Rilis} experiment. As seen, the values obtained for
successive cycles fluctuate more than statistics allow, leading to
large values of $\chi$. This is due to the short term fluctuations
of the magnetic field. These fluctuations are taken into account by
adding a `static' instrumental error, $\sigma_{st}$, to each cycle.
The averaged $\overline{\chi}$ is then reduced from 1.5 to 1 as
shown in Fig. \ref{fig:chi4}-b (further discussions are given in
Section \,\ref{sec:mgcalib}). This error is of the order of a few
$10^{-7}$.

When the same mass measurement is repeated after changing some
settings, the measured values may eventually be more dispersed than
expected, requiring the introduction of a `dynamic' instrumental
error, $\sigma_{dyn}$, determined so as to reduce $\chi$ to unity.

At each step of the analysis, the implied errors are quadratically
combined.

From the measured ion mass $m_x$, we deduce the atomic mass
$M_x^{meas}$, after taking into account the relativistic velocity
correction. In order to make the results easier to compare, they are
expressed not in mass values but rather in relative mass
differences:
   \begin{equation}    \label{eq:MDRE}
    \Delta_x^{meas} = (M_x^{meas} - M_x^{table})/ M_x^{table}\;.
   \end{equation}
   The $M_x^{table}$ values are arbitrary test values used to define
differences. In the final results they will be added back to our
values so that they do not affect the result. They have as such no
associated errors. For convenience we have adopted for this role the
unrounded mass values from the 1995 files of {\sc Ame'95}
\cite{Ame95bis}. The results of the three experiments are compared
with the {\sc Ame'95} \cite{Ame95} and not the {\sc Ame'2003}
\cite{Ame03} which includes already our results, some of which were
preliminary.

Before presenting the results (sections
\ref{sec:rilis},\ref{sec:plasma}, and \ref{sec:thermo}), we discuss
one more important source of error: calibration of the reference and
measured masses.

\section{Calibration}  \label{sec:calib}

f the superposition of the two beams in the magnetic field is
imperfect, at a given point their trajectories will deviate from
each other by $\delta l$. If, at this point, the magnetic field is
not homogeneous, the ions will then experience a different magnetic
field, according to
   \begin{equation}    \label{eq:traj}
   \frac{\delta B}{B}=\frac{1}{B}\frac{dB}{dl}\delta l\;.
   \end{equation}
A mass shift, $\Delta_x^{meas}$, will consequently be observed,
proportional to the integral of Eq. (\ref{eq:traj}) along the
trajectory between the two modulations. Since the gradients,
$\frac{1}{B}\frac{dB}{dl}$, measured in the {\sc Mistral} magnetic
field were of the order of 20-50$\times10^{-7}$/mm \cite{coc91}, a
few millimeters displacement of the trajectory may be enough to
produce shifts which may be of the order of 10$^{-5}$ when the
reference and measured masses are very different.

The deviation of the trajectory can have various origins : ($i$) the
high voltage power supplies for $V_r$ and $V_x$ in Eq.~(\ref{eq:mv})
are not perfectly calibrated, ($ii$) the delay time required for the
voltage jump is not sufficient, or ($iii$) the reference ion beam is
not emitted in the same way when $V_r$ is changed (the voltage of
{\sc Isolde} remains unchanged during the run). Furthermore a shift
may also occur between beams from {\sc Isolde} and the reference ion
of the same mass. So the relation connecting $\delta l$ to
$(V_r,V_x)$ or $(m_r,m_x)$ remains unknown. In our earlier work
\cite{Lun01a}, we tested various scenarios, to characterize the jump
amplitude, which gave very similar results. We therefore adopted
there the simplest one
   \begin{equation} \label{eq:MDRE_old}
   \Delta_x^{meas} = a(m_r - m_x) + b\;,
   \end{equation}
where the constant term $b$ takes into account the offset which is
observed between the {\sc Isolde} and {\sc Mistral} beams of the
same mass (for example $^{14}$N$^{14}$N and $^{28}$Mg in the {\sc
Rilis} experiment). However taking into account all the results
presented here and some more recent ones on Li and Be isotopes
\cite{Bac04}, it now appears that a calibration law of the form
   \begin{equation}    \label{eq:calib}
   \Delta_x^{meas} = a \frac{\Delta V}{\overline{V}} + b = a \frac{\Delta (m/q)}{\overline{(m/q)}} +
   b\;,
   \end{equation}
with $\Delta V = V_x - V_r$, $\overline{V}=(V_x + V_r)/2$, $\Delta
(m/q) = (m_r/q_r) - (m_x/q_x)$, and \mbox{$\overline{(m/q)} =
[(m_r/q_r) + (m_x/q_x)]/2$}, fits better. It is also more satisfying
since it only includes relative variations, and is compatible with
assumptions ($i$) and ($iii$). Assumption ($ii$) leads to a more
sophisticated dependance on $V_x$ and $V_r$. In any case, we imposed
a sufficient delay to avoid that condition.

If the relative difference $\Delta V/\overline{V}$ is large, it
would seem better to consider Eq. (\ref{eq:calib}) as a differential
equation and to integrate it:
   \begin{equation}
   \label{eq:calib_ln}
   \Delta_x^{meas} = a \ln (V_x/V_r) +b \; .
   \end{equation}

Both formulae (\ref{eq:calib}) and (\ref{eq:calib_ln}) have the same
first- and second-order terms, so that they are equivalent as long
as $\Delta V/\overline{V}$ is less than about 30 \%. In most cases
this condition was fulfilled, and even when it was not, the induced
variations on the final results were completely negligible compared
to the assigned errors.

As mentioned in \cite{Lun01a} §IIIA, in cases where the same mass
$m_x$ is compared to two reference masses $m_{r1}$ and $m_{r2}$
without any changes in the settings of the $m_x$ beam, the
difference between the two measurements, $\Delta_{x,r1}^{meas} -
\Delta_{x,r2}^{meas}$, may be used for calibration even if the value
of $m_x$ is unknown, since it is equivalent to a direct comparison
of the masses  $m_{r1}$ and $m_{r2}$. From a mathematical point of
view, Eq.~(\ref{eq:calib_ln}) is coherent with this assertion, while
it is not the case for Eq.~(\ref{eq:calib}) since we have:
\begin{equation}
\frac{V_{r2}-V_{r1}}{V_{r2}+V_{r1}}\neq
\frac{V_x-V_{r1}}{V_x+V_{r1}} - \frac{V_x-V_{r2}}{V_x+V_{r2}}
.\end{equation} However this difference induces only very small
changes ($< 10^{-7}$) in the present experiments.

The calibration was performed using both relations (\ref{eq:calib})
and (\ref{eq:calib_ln}), and we found as expected that the
differences between the final results are negligible. So finally, we
made the choice to present  ~here calibration laws given by
Eq.(\ref{eq:calib}) which is more intuitive.

\begin{figure}
   \begin{center}
   \includegraphics[width=20pc]{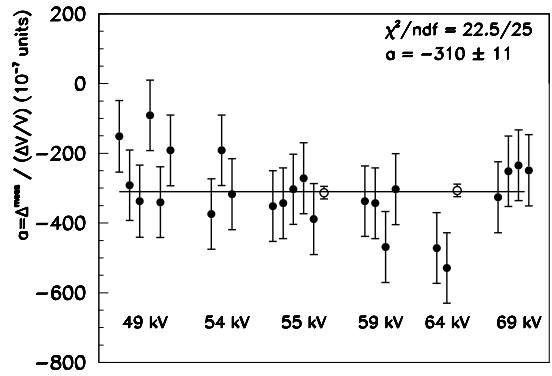}
   \caption{\small
   Calibration slopes $a=\frac{\Delta_x^{meas}}{\frac{\Delta V}{\overline{V}}}$ obtained for $^{14}$N$^{14}$N $-
^{15}$N$^{14}$N (full circles) or $^{14}$N$^{14}$N $- ^{23}$Na
(empty cirles) for various accelerating voltages $\overline{V}$. }
   \label{fig:caltst}
   \end{center}
   \end{figure}

  An important point is that the calibration law implies that the same mass difference measured at
different accelerating voltages will lead to the same slope $a$.
   The comparison of the masses of the molecules $^{14}$N$^{14}$N and
$^{15}$N$^{14}$N was performed for accelerating voltages varying
from 50 to 70~kV.
   Two measurements were also performed comparing $^{14}$N$^{14}$N and
$^{23}$Na.
   Figure \ref{fig:caltst} shows the results obtained after adding
quadratically a `{dynamic}' error,  $\sigma_{dyn} = 3.5 \times
10^{-7}$, to each measurement.
   This quantity leads to a good consistency ($\chi^2$~=~0.90) of all the
measurements.

The parameters $a$ and $b$ are determined by a least-squares fit
using `calibrant' masses. The `calibrant' masses are those known
with an accuracy better than 10$^{-7}$, and determined by at least
two different methods with agreeing results. The values of
$\Delta^{meas}_{x}$ can then be corrected using the calibration law:
   \begin{equation}    \label{eq:corr}
   \Delta^{corr}_{x} =
     \Delta^{meas}_{x}
   - \left(a \frac{\Delta V}{\overline{V}}
   + b \right)\;.
   \end{equation}
This value must be, of course, compatible with zero when calculated
for `calibrants'.

The value of the corrected relative mass difference is also
extracted for the measured masses according to relation
(\ref{eq:corr}). The error of $\Delta^{corr}_{x}$ for measured
masses is a quadratic combination of the statistical, the `static',
the `dynamic' and the `calibration' errors. The `calibration' error
is given by
   \begin{equation}    \label{eq:errcal}
   \sigma^{2}_{cal} =
   \left(\frac{\Delta V}{\overline{V}}\right)^2 \sigma _a^2
   + \sigma^2 _ b
   + 2 \left(\frac{\Delta V}{\overline{V}}\right) \sigma _{ab} \; ,
   \end{equation}
where $\sigma_a$, $\sigma_b$ and $\sigma_{ab}$ are determined from
the least-squares fit of the calibration law.

   \section{Analysis of the {\sc Rilis} experiment}
   \label{sec:rilis}

    The resonant laser ionization ({\sc Rilis}) mode was chosen in
order to selectively ionize magnesium isotopes. Alkali elements (Na)
are still present due to thermo-ionization, but they are much less
produced than the Mg isobars.

   \subsection{Calibration of the {\sc Rilis} experiment}
   \label{sec:mgcalib}

   It is desirable to choose the reference masses in such a way as to
minimize the amplitude of the jumps. The reference mass was that of
the molecule $^{14}$N$^{14}$N at $A=28$.
   The quantities $a$ and $b$ (Eq. (\ref{eq:calib})) have been determined by measuring the precisely
known `calibrant' masses: $^{23}$Na and $^{24-26,28}$Mg.
   To decrease the contributions of the `calibration' error, we
interspersed measurements of calibration masses and of unknown masses.

\begin{figure}
   \begin{center}
   \includegraphics[width=19pc] {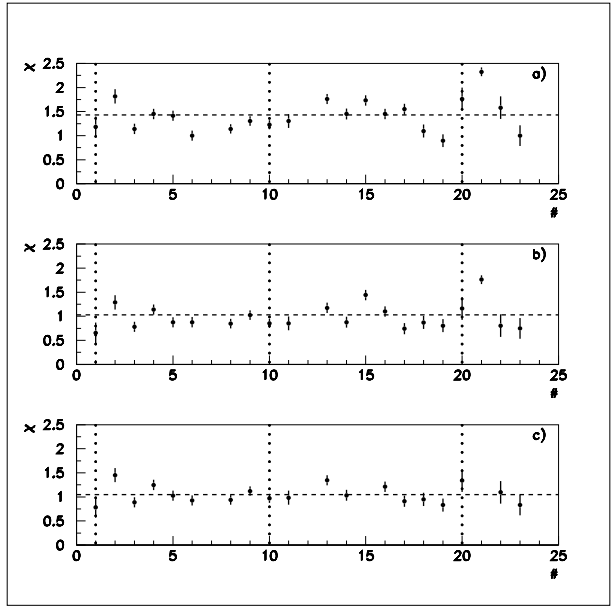}
   \caption{\small
   The $\chi$ values for each of the 23 mass measurements performed during the {\sc Rilis} experiment in
chronological order
   (a) before adding any `{static}' instrumental error ($\overline{\chi} = 1.4$);
   (b) after adding a `{static}' instrumental error of $4 \times 10^{-7}$ ($\overline{\chi} = 1.0$);
   (c) after removing measurements 15 and 21 and reducing the `{static}' instrumental error
to $3 \times 10^{-7}$ ($\overline{\chi} = 1.0$).
   }
   \label{fig:chi4}
   \end{center}
   \end{figure}

  The $\chi$-value of the average mass
for each measurement (that is 21 measurements out of 23) was plotted
in Fig.\,\ref{fig:chi4}-a.
   The two measurements (number 7 and 12) not represented, correspond to $^{32,33}$Mg which have been
analyzed differently due to lower statistics (several cycles had to
be summed).
   In order to restore the normalized $\chi$ for the average to unity,
we added a `{static}' error, $\sigma_{st} = 4 \times 10^{-7}$, to
each cycle (see Fig.\,\ref{fig:chi4}-b). However, measurements 15
and 21 still had high $\chi$-values. Examining the dispersion within
each measurement, we also found out that for these two measurements
the results were more dispersed ($11 \times 10^{-7}$ instead of
$\sim 6 \times 10^{-7}$ on average). Therefore it was decided to
remove measurements 15 and 21.
   As a consequence, the `{static}' error required to restore the
Birge ratio to unity is reduced from 4 to $3 \times 10^{-7}$ (see
Fig.\,\ref{fig:chi4}-c).

To get an estimate of the `{dynamic}' error, 6 measurements
($^{23}$Na, $^{24-26,30-31}$Mg) were repeated 2 or 3 times. The
observed dispersion is 3.8$\times 10^{-7}$ leading to
\mbox{$\sigma_{dyn} = 4 \times 10^{-7}$}.

\begin{figure}
\begin{center}
   \includegraphics[width=20pc]{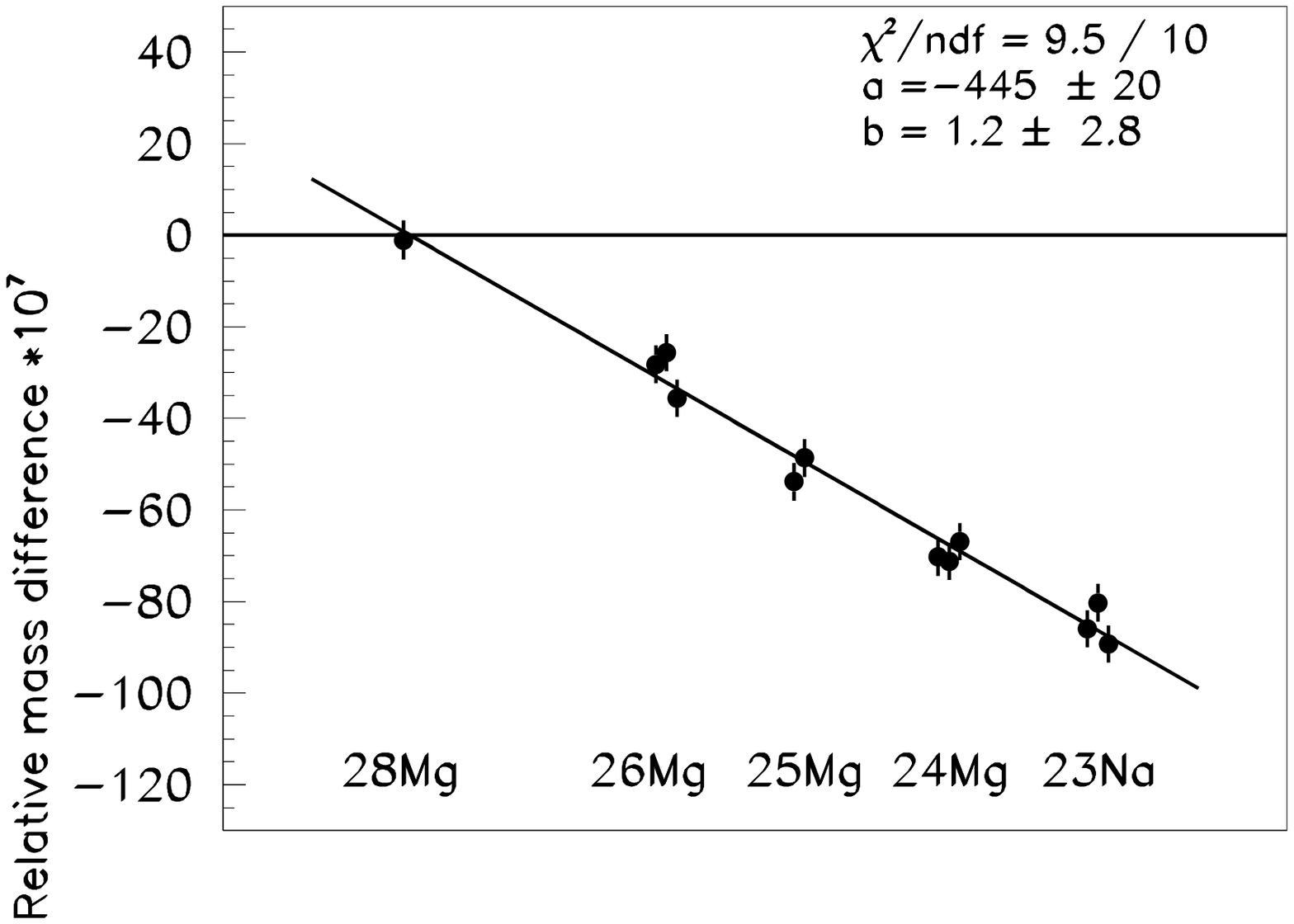}
\vskip -1.72cm
   \includegraphics[width=20pc]{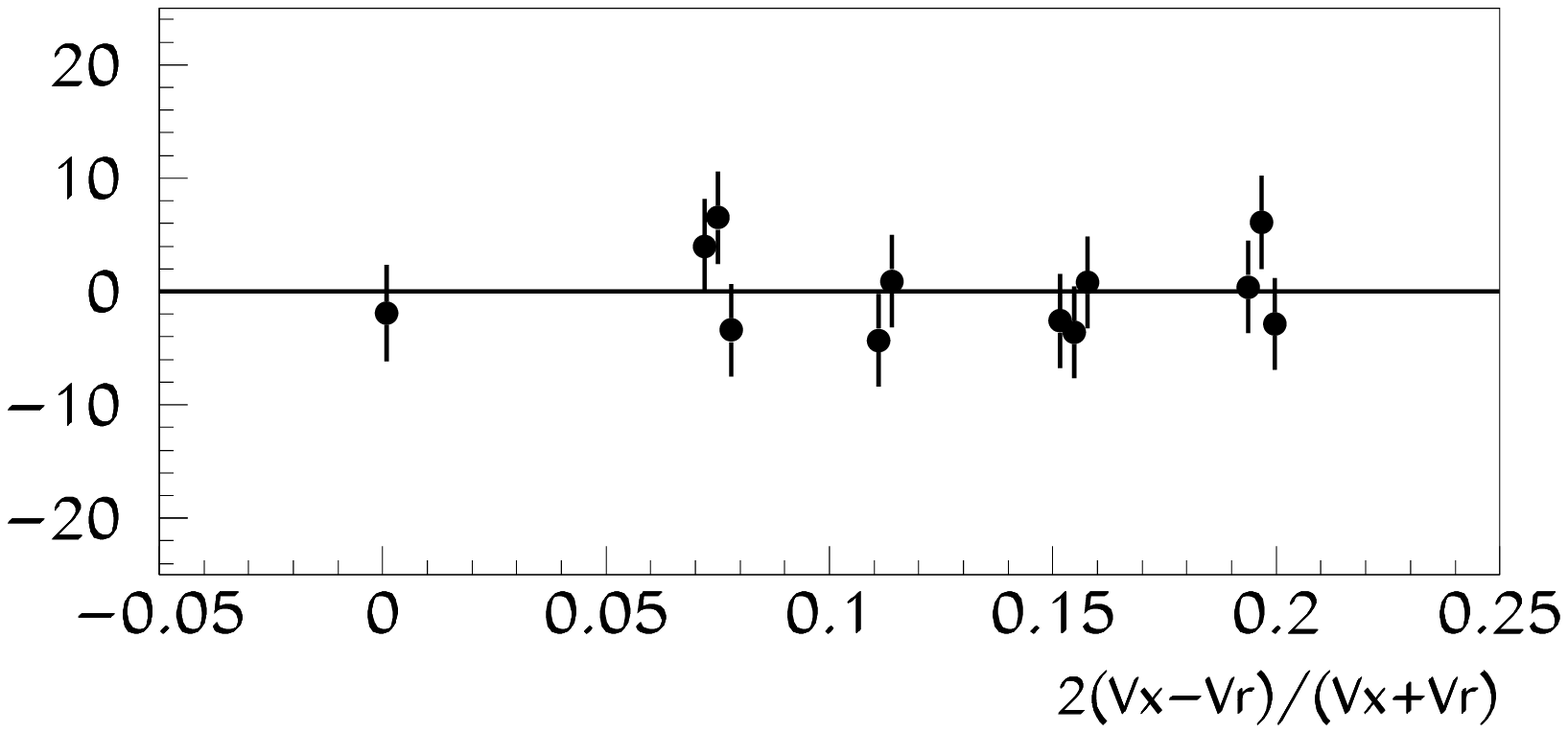}
   \caption{\small
   Calibration function for the {\sc Rilis} experiment for all calibrant masses.
   Top: plotted is the relative mass differences $\Delta^{meas}_{x}$, for the calibrant masses,
   versus the relative voltage difference
   $\Delta V / \overline{V}$.
   The continuous line represents the fit for the linear calibration
   law (Eq.~(\ref{eq:calib})). Bottom: residual after subtraction of the calibration law.
   }
   \label{fig:CalMg}
\end{center}
\end{figure}

By fitting together the twelve calibration measurements listed in
Table\,\ref{tab:CalMg}, we were able to determine the parameters,
$a$ (slope) and $b$ (offset), of the calibration law
(Eq.~(\ref{eq:calib})): $a=-445(20)\times10^{-7}$,
\mbox{$b=1.2(2.8)\times10^{-7}$}, and a correlation coefficient
$\sigma^2_{ab}=-53\times10^{-14}$. The $\chi^2$ value is 0.95, which
shows that the data are in good agreement with the chosen
calibration law, as shown in Fig.~\ref{fig:CalMg}.

   The complete set of measured relative mass differences of calibrants is shown
in Table\,\ref{tab:CalMg} column~2. The resulting relative mass
differences are presented individually in the first part of Table
\ref{tab:CalMg}, and their averaged values are given in the second
part. All $\Delta_x^{corr}$ values are compatible with zero, as
expected.

\subsection{Results of the {\sc Rilis} experiment}
\label{sec:MgRes}

New precise measurements on $^{26}$Na and $^{29-33}$Mg have been
achieved. An example of a recorded mass peak for $^{33}$Mg was shown
in Fig.\,\ref{fig:33mgpic}. The final results of the present {\sc
Mistral} experiment, expressed in $\mu$u, are reported in column 5
of Table\,\ref{tab:ResRilis}.

\begin{figure}
   \begin{center}
   \includegraphics[width=25pc]{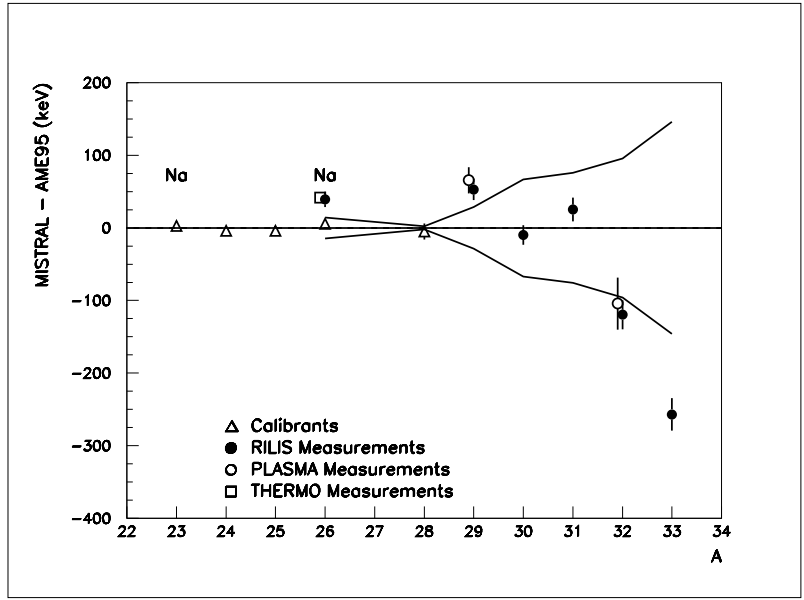}
   \caption[]{\small
   Comparison of the {\sc Rilis} measurements with the
previously adopted mass values from {\sc Ame'95} \cite{Ame95}. The
error is a quadratic combination of all sources of errors. The {\sc
Ame'95} precision is represented by the two symmetric continuous
lines. The {\sc Plasma} results for $^{29,32}$Mg and those of the
{\sc Thermo} experiment for the $^{26}$Na are also reported here for
comparison.
   }\label{fig:ResMg}
   \end{center}
   \end{figure}

Figure \ref{fig:ResMg} shows that the {\sc Ame'95} recommended
values are in rather good agreement for $^{30,31,32}$Mg while nearly
a $2\sigma$ deviation is observed for $^{29}$Mg and $^{33}$Mg, and
$3\sigma$ for $^{26}$Na. The Mg discrepancies are discussed in
\cite{Lun04}. One can notice that the precision is improved by a
factor 2 to 7 as compared to {\sc Ame'95}. The other interesting
point is that $^{32,33}$Mg ($N=20,21$) are found more bound by
respectively 120 and 260~keV, reinforcing even further the collapse
of the $N=20$ shell closure as discussed in~\cite{Lun04}.

\section{Reanalysis of the {\sc Plasma} experiment}
\label{sec:plasma}

Neutron-rich Mg and Ne isotopes were the subject of an earlier
experiment (1999), for which very preliminary results appeared in
the literature \cite {Lun01b}.

\subsection{Calibration of the {\sc Plasma} experiment}
\label{sec:plCalib}

At {\sc Isolde}, a plasma ion source was used, which provided many
elements but also molecules and doubly-ionized ions. The advantage
was to have many isobars for calibration, but avoiding contamination
was a challenge since {\sc Mistral} produces transmission peaks for
every ion species and every harmonic number. Thus, even an isobar
with a relatively large mass difference can interfere with the mass
of interest. Two unfortunate features of this experiment must also
be mentioned : many readjustments of the ion optics were necessary
during the experiment, and the insulators of the injection device
were permanently discharging.

Three reference masses from {\sc Mistral} were used: $^{23}$Na for
measurements on masses 23-29, and $^{39,41}$K for measurements on
masses 30-41.

A first analysis was performed \cite{Mon00,Lun01b} using the old
calibration law , Eq.~(\ref{eq:MDRE_old}). There was no way to fit a
unique calibration law for the two subsets of data relative to the
 $^{23}$Na and $^{39,41}$K  references. For the
data taken with Na as reference, the calibrants were $^{23}$Ne,
$^{23,25}$Na, $^{25-27,29}$Mg and $^{27,29}$Al. A calibration law
could be fitted with the addition of a time-dependant linear term
for the offset ($a \Delta m + b + c\times t$). The `dynamic' error
was found $\sigma_{dyn} = 12\times 10^{-7}$ . Mass values were
deduced for $^{25,26}$Ne, while $^{29}$Mg which was used as a
calibrant, was deviating slightly. However, for the data using
$^{39,41}$K as references, the situation was very confused: there
was a lack of calibrants, and the data not only were at variance
with the $^{23}$Na reference ones but also seemed hardly consistent
with each other. A mass value was finally given for  $^{32}$Mg with
an adopted error taking into account these uncertainties.

A completely new analysis of these data has now been performed. To
start with, special attention was devoted to identify the various
recorded peaks so that the number of calibrants was increased,
especially for the set using the $^{39,41}$K reference:
singly-charged $^{29}$Si,
 $^{13}$C$^{16}$O, $^{14}$N$^{15}$N, $^{12}$C$^{18}$O,
$^{15}$N$^{15}$N, $^{32}$S ions, and doubly-charged $^{60}$Ni,
$^{78,82}$Kr ions were identified. \textit{A contrario}, $^{29}$Mg
was considered as an unknown mass to be determined, and some
low-quality measurements were rejected. The individual results for
the calibrants are given in Table \ref{tab:CalMg99}.

As explained above, the first step of this analysis aims to
determine the `static' and `dynamic' errors. The `static' error is
found to be $\sigma_{st} = 5\times 10^{-7}$. Before determining the
`dynamic' error, a time-dependant linear term very similar to the
one of the old analysis, had to be adjusted. The reproducibility of
the results was examined for each nuclear species (39 values
distributed between 11 series) as well as for each jump amplitude
(44 values distributed between 9 series). Both studies lead to a
time constant $c=0.5\times 10^{-9}$~/min and $\sigma_{dyn} = 7\times
10^{-7}$, which is already larger than that observed in the {\sc
Rilis} experiment.

Once this time-dependant term is subtracted, the second step is to
fit the calibration law. The offsets were found to depend on the
reference (Na or K). The fitted calibration law is thus :
   \begin{equation} \label{eq:plcalib1}
   \Delta_x{}_t^{meas} = \Delta_x^{meas} - 0.5\times 10^{-9} t_{min} = a \frac{\Delta V}{\overline{V}} +b_{Na}
   \end{equation}
for $^{23}$Na reference (32 data), and
   \begin{equation} \label{eq:plcalib2}
   \Delta_x{}_t^{meas} = \Delta_x^{meas} - 0.5\times 10^{-9} t_{min} = a \frac{\Delta V}{\overline{V}} +b_{K}
   \end{equation}
 for $^{39,41}$K references (13 data).

\begin{figure}
   \begin{center}
   \includegraphics[width=30pc] {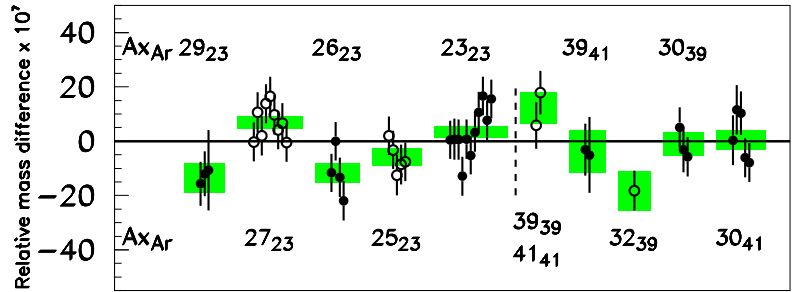}
   \caption{\small
  Dispersion of the measurements corresponding to the same jump amplitude, as compared to the fitted calibration law after
  addition of $\sigma_{dyn} = 7\times 10^{-7}$. A time dependance of
  $0.5\times 10^{-9}$/min has been subtracted before averaging. The
  shadowed areas correspond to the 1$\sigma$ limit of the averaged
  values of each group.
   }
   \label{fig:plasmadisp}
   \end{center}
   \end{figure}

This linear calibration fit leads to $\chi^2/ndf = 77/42$ which is
unsatisfactory. Furthermore, if all measurements corresponding to
the same jump amplitude are averaged before the fit, we obtain
$\chi^2/ndf = 41.4/7$, as shown in Fig. \ref{fig:plasmadisp}. This
result reveals the presence of strong fluctuations depending on the
applied voltages, which are thought to be mainly related to the
insulator discharges. In order to incorporate these unexpected
fluctuations, the weighted mean values calculated for each jump (10
data) have been used as input values for the linear calibration fit
instead of the individual ones. A $\chi^2$ close to unity was
obtained by adding a `fitting' error $\sigma_{fit} = 9\times
10^{-7}$  to each input value.

\begin{figure}
\begin{center}
   \includegraphics[width=20pc]{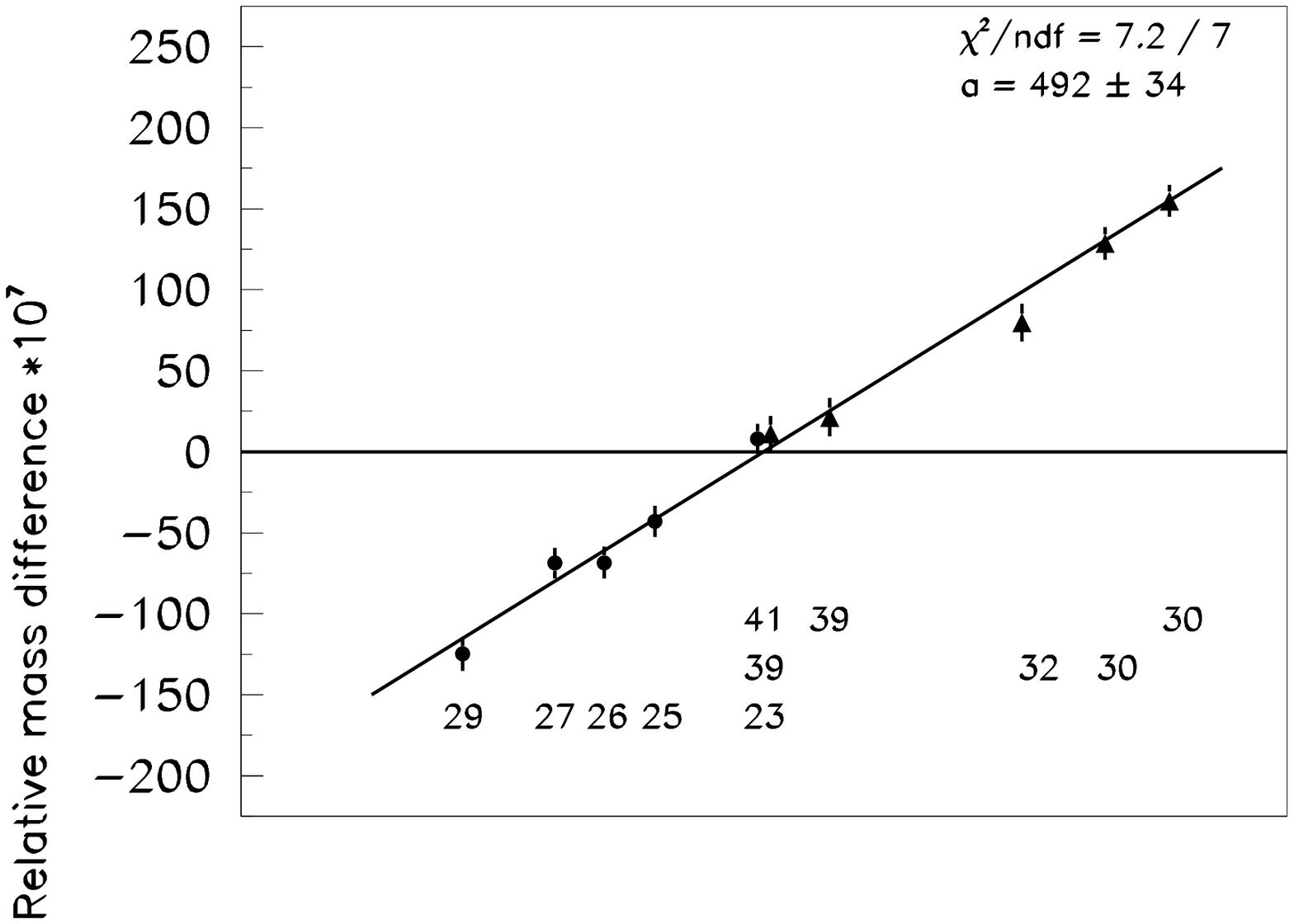}
\vskip -1.72cm
   \includegraphics[width=20pc]{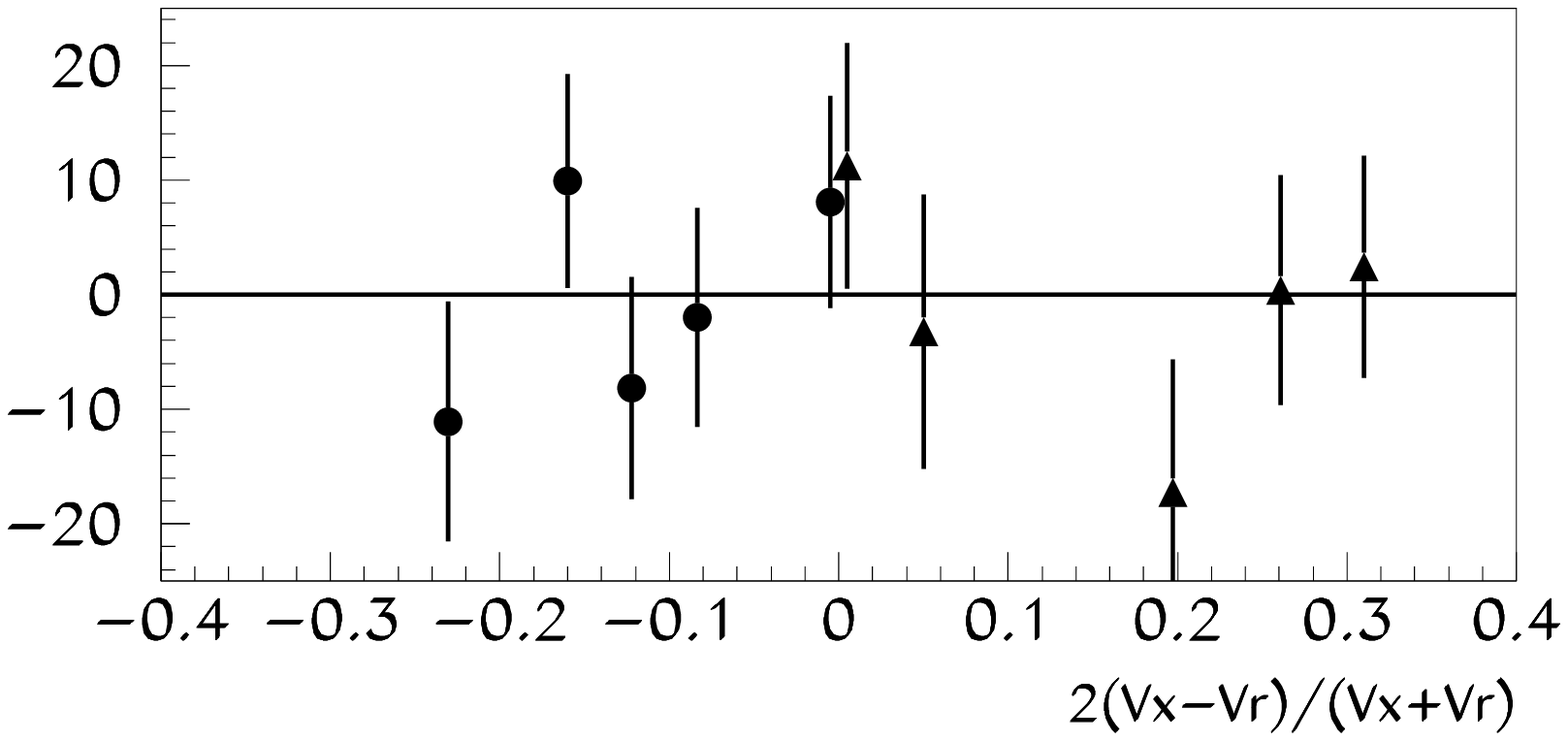}
   \caption{\small Calibration function for the {\sc Plasma}
   experiment for all calibrants taken together.
 Plotted are the weighted mean values  of the relative mass
 differences $\Delta^{meas}_{x}$  corresponding to the same jump
 amplitude
versus the relative voltage differences $\Delta V / \overline{V}$.
   The $A$ values
   are indicated for each input data (the references appear at
   $\frac{\Delta V}{\overline V}=0$). The continuous lines represent
   the fit for the linear calibration law (Eqs.\,(\ref{eq:plcalib1}) and (\ref{eq:plcalib2})).
Top: after subtraction of the offsets. Bottom: after subtraction of
the full calibration law. The full circles are for the Na reference,
and the triangles for the K reference.} \label{fig:plasmacalib}
\end{center}
\end{figure}

The fitted parameters and their uncertainties are
 $a=492(34)\times10^{-7}$,
\mbox{$b_{Na}=6.4(5.8)\times10^{-7}$},
\mbox{$b_{K}=-24.6(8.7)\times10^{-7}$}, and the correlation
coefficients are $\sigma^2_{ab(Na)}=136\times10^{-14}$ and
$\sigma^2_{ab (K)}=-238\times10^{-14}$. Table \ref{tab:plasmacalib}
gives the values of $\Delta_x^{meas}$ and $\Delta_x^{corr}$ for the
averaged masses of the calibrants. The deviations, in column 4, from
the calibration law, are compatible with zero within the error bars,
as expected. Figure~\ref{fig:plasmacalib} illustrates the result of
this fit. It is clearly seen that the different voltages generate
fluctuations around the linear calibration law and not a systematic
deviation from it.

A fit performed on the same data using $\Delta m$ instead of $\Delta
V/\overline V$ gives $\chi^2=2.4$ instead of 1.0, which clearly
demonstrates the superiority of the new calibration law.

\subsection{Results of the {\sc Plasma} experiment}  \label{sec:plres}

Using the calibration law obtained in section \ref{sec:plCalib}, we
obtain mass values for $^{26}$Ne and $^{29,32}$Mg. Concerning
$^{25}$Ne, which was mentioned in references~\cite{Mon00,Lun01b}, it
appeared that the peak was an unreliable double peak so this
measurement is to be retracted. Concerning $^{29}$Mg which had been
used as a calibrant while its mass precision, from the {\sc Ame'95}
\cite{Ame95}, was rather low, it has now been considered as a
measurement. The obtained values are given in
Table~\ref{tab:mg99-res1}. The preliminary values
\cite{Mon00,Lun01b} differed from the present ones by $-19(33)$
$\mu$u, $20(26)$ $\mu$u, and $-176(113)$ $\mu$u for $^{26}$Ne,
$^{29}$Mg, $^{32}$Mg, respectively.

As noticed in section \ref{sec:plCalib}, Fig.~\ref{fig:plasmacalib}
exhibits strong fluctuations depending on the mass jump while
Fig.~\ref{fig:plasmadisp} demonstrates a good coherence  between
data corresponding to the same jump amplitude. It thus appears that
more reliable results can be derived when direct comparison between
the measured values and the weighted average of the isobaric
measurements is performed, as it is possible for $^{26}$Ne and
$^{29}$Mg. In order to be free of the time dependance of the offset,
only data taken during a small period of time have been considered.
However, some settings may have changed, so that a `dynamic' error
is still to be considered, but appears to be slightly improved (from
7 to 6 $\times 10^{-7}$). The obtained values are given in
Table~\ref{tab:mg99-res2}. In the case of $^{26}$Ne, only the two
$^{26}$Mg measurements performed at time 5430 and 5520 (see Table
\ref{tab:CalMg99}) obey this constraint. This procedure is confirmed
by three measurements where we compared directly the masses of two
calibrant isobars produced by {\sc Isolde} :
\begin{eqnarray*}
^{25}\textrm{Mg}-^{25}\textrm{Na} & : & \Delta^{meas} \times 10^7 = -2.7(2.3)\\
^{27}\textrm{Mg}-^{27}\textrm{Al} & : & \Delta^{meas} \times 10^7 = 1.4(3.2)\\
^{27}\textrm{Mg}-^{27}\textrm{Al} & : & \Delta^{meas} \times 10^7 =
3.9(2.2)
\end{eqnarray*}
The values of Table \ref{tab:mg99-res2} are compatible with those
obtained with the calibration law, but their uncertainties are much
lower. We indeed adopt $^{26}$Ne and $^{29}$Mg from the isobaric set
and $^{32}$Mg from the average as given in Table
\ref{tab:mg99-res3}. They agree with the results of the {\sc Rilis}
experiment. These results supersede those of the preliminary
analysis ~\cite{Mon00,Lun01b}.

   \section{Reanalysis of the {\sc Thermo} experiment}
\label{sec:thermo}

Neutron-rich Na isotopes were the subject of an earlier experiment
(1998), for which the results are reported in reference
\cite{Lun01a}.

   \subsection{Calibration of the {\sc Thermo} experiment}  \label{sec:naCalib}

In the Na measurements, the calibration law was the one given in Eq.
(\ref{eq:MDRE_old}).
   Various laws, including the present one, had been tested at that time
without any noticeable consequence in the results.
   However, for consistency, and since we now have a better knowledge
of the calibration law, we have reanalyzed the sodium measurements,
using the same procedure as used here for the {\sc Rilis}
experiment.

In reference \cite{Lun01a}, the concepts of `static', `dynamic' and
`fitting' errors were not introduced. Instead a `systematic' error
of $5 \times 10^{-7}$ was found to achieve consistency of the data
with the calibration law. The static error may be neglected. As the
dynamic error cannot be estimated due to the lack of data available
for this task, a $4\times 10^{-7}$ uncertainty is directly adjusted
and sufficient to fit the new calibration law. The new calibration
parameters are given in Table \ref{tab:newfitNa} which supersedes
Table II published in \cite{Lun01a}. (Note that $a$ no longer has
the same units.)

   The resulting relative mass differences are presented individually
in the first part of Table \ref{tab:CalNa}, and their averaged
values are given in the second part. This table supersedes Tables
I+III+V (concerning the calibration results) published in
\cite{Lun01a}. As expected, these values are very close to the
previously published ones.

   The general agreement is quite good: the global values of $\chi$ for
the calibrants is 0.6.
   Comparing the new values to the published ones, it appears that the
main changes occur in set 1.
   This is quite understandable since the slope $a$ has a large value as
compared to the slopes obtained in set 2.
   However, the changes do not exceed half a standard deviation, confirming that the
particular choice of calibration procedure was not crucial in that
experiment.

\subsection{Results of the {\sc Thermo} experiment}
\label{sec:NaRes}

  The updated final results are given individually in Table~\ref{tab:ResNa1}
and averaged in Table~\ref{tab:ResNa2}. These tables, respectively,
supersede Tables I+III (concerning the measured masses) and VI
published in \cite{Lun01a}.
   The general agreement is quite good: the global values of $\chi$ for
the measurements is 1.0.

The averaged value of $\Delta^{corr}_{x} \times 10^7$ obtained for
$^{27}$Na is 24.8(3.1). However, an isobaric measurement analog to
that described in the {\sc plasma} experiment could already be
performed in \cite{Lun01a}. This result being more precise, it is
the one reported in Table~\ref{tab:ResNa2}.

\section{Data evaluation}

All the new reevaluated results are summarized in
Table~\ref{tab:summ}.

In the three experiments described here and all using the {\sc
Mistral} spectrometer, the mass determination was repeated for three
nuclides:
 $^{26}$Na in the {\sc Thermo} and {\sc Rilis} experiments;
$^{29}$Mg and $^{32}$Mg in the {\sc Plasma} and {\sc Rilis}
experiments. Table~\ref{tab:summ} shows for these three repeated
mass determinations a very good agreement confirming that the errors
evaluations and assignments have been correctly established.

The next step, in order to prove accuracy, is to establish that no
residual errors appear when comparing to previously known masses. We
therefore compare our results to those of other experiments where
they exist and when they were not used as calibrants.

   Such comparison and discussion for $^{29-33}$Mg is done in \cite{Lun04}.
   It is shown there that there is no serious
    conflict with other measurements. A similar study for
    the results of the {\sc Thermo} experiment is developed in
    \cite{Lun01a} and its conclusions are still
    valid.
    We therefore shall limit here the discussion to the results of $^{26}$Ne and $^{26}$Na.

   Previously, the mass of $^{26}$Ne has been determined in two very
different types of experiment: a charge-exchange reaction
$^{26}$Mg($\pi^-$,$\pi^+$)$^{26}$Ne \cite{80Na12} reported a
$Q$-value of $-17\,676(72)$\,keV corresponding to a $^{26}$Ne mass
excess of 472(77)\,$\mu$u;
   and
   a time-of-flight (TOF) experiment reported in 1991 \cite{91Or01}
yielded a value 448(90)\,$\mu$u for $^{26}$Ne.
   The agreement with our value, 518(20)\,$\mu$u, is excellent.

   The mass of $^{26}$Na has been determined in several ways.
All our data from Table~\ref{tab:summ} have been inserted in the
atomic mass evaluation and a new adjustment has been performed.
Table~\ref{tab:disc} gives all data related to $^{26}$Na for which
the precision of the new adjusted value is better than 25\,$\mu$u:
\begin{itemize}
    \item A $\beta^-$-decay energy measurement \cite{73Al13} reported in 1973 a
$Q_{\beta}$-value of 9\,210(200)\,keV, which fully agrees with
$Q_{\beta}=9\,354(4)$\,keV deduced from our result.

    \item A ($^7$Li,$^7$Be) reaction on $^{26}$Mg at Chalk-River in 1972
\cite{72Ba35} yielded an energy of $Q=-10\,222(30)$\,keV that was
corrected in the {\sc Ame2003} to $-10\,182(40)$\,keV to take into
account the contribution of the unresolved 82.5\,keV level.
   Effectively, the peak corresponding to the $^{26}$Na ground-state
appears wider (Fig.\,2b in \cite{72Ba35}).
   The corresponding value derived from our result for $^{26}$Na is
$-10\,216(4)$\,keV, in good agreement (0.8\,$\sigma$) with the
corrected value, but also with the original one.

    \item In another charge-exchange reaction using a tritium projectile
$^{26}$Mg($^3$H,$^3$He)$^{26}$Na performed at Los Alamos
\cite{74Fl01}, an energy $Q=-9\,292(20)$\,keV was obtained.
   Our result for $^{26}$Na does not agree with \cite{74Fl01} at a
$2\sigma$ level, our value implying $Q=-9\,335(4)$\,keV. The result
of \cite{74Fl01} has been obtained with a good population of the
peaks and with enough resolving power to be able to detect for the
first time the peak at $82.5$\,keV from a state with half-life $\sim
9 \mu$s.
   An explanation could be that the calibration of their measurement,
against the $^{16}$O($^3$H,$^3$He)$^{16}$N reaction, provides energy
knowledge in a region of the spectra far from where the $^{26}$Na
ground-state peak occurs (see Fig.\,1 in \cite{74Fl01}).
   It would be interesting to deduce the ground-state mass excess from
the determination of the masses of the excited states at 1\,996 and
2\,048\,keV.
   In any case, we consider the measurement of Los Alamos as the most
important conflict to our result.
   It is hoped that a measurement of the mass of $^{26}$Na with a
Penning trap would give a clear answer to this problem.

    \item Finally, a series of mass-spectrometric measurements involving
$^{26}$Na was performed by our Orsay-group at {\sc Cern} in the
pioneering work of on-line mass-spectrometry, in the early 1970's
\cite{75Th08}.
   From the 16 measurements selected in Table~\ref{tab:disc}, one can see that two
results are at 1.8$\sigma$ from our present knowledge.
   Statistically these numbers are acceptable.
We consider therefore that no serious conflict exists with
\cite{75Th08}.
\end{itemize}

The overall good consistency of our results with others where they
exist, as discussed above and in \cite{Lun04}, and \cite{Lun01a},
points out accuracy of {\sc Mistral} measurements.

   \section{Conclusion}

   We have presented here a new very careful analysis of three long
   series of mass measurements performed by the {\sc Mistral}
   spectrometer, with the aim to evaluate all sources of errors and
   to look for the best calibration law. For these three
experiments, the values of the measured `calibrant' masses are
compatible with their precisely known values from {\sc AME'95}
\cite{Ame95}, as expected.

The previously measured masses of $^{26}$Ne and $^{29,32}$Mg ({\sc
Plasma} experiment) and of $^{26-30}$Na ({\sc Thermo} experiment)
have been reevaluated with respect to the newly established
calibration law. For the {\sc Thermo} experiment, the changes do not
exceed half a standard deviation, confirming that the particular
choice of a calibration procedure was not crucial in that
experiment. However, the {\sc Plasma} experiment demonstrates
clearly the superiority of this new calibration law.

The newly measured mass values obtained for the series of isotopes
$^{30-33}$Mg shows that the {\sc Mistral} spectrometer can achieve
high accuracy even for short lived nuclides. The results obtained in
the {\sc Rilis} experiment have a precision 5 to 7 times better than
in the existing literature \cite{Ame95}. The detailed evaluation of
these nuclides can be found in \cite{Lun04}. It is worthwhile to
stress on the impact of this new precise and accurate mass of
$^{33}$Mg, since it is used as a calibrant in the SPEG experiment at
GANIL \cite{sar00}, and it moved considerably (276 $\mu$u).

Due to the rejection of the preliminary result of the $^{25}$Ne in
the {\sc Plasma} experiment (see section \ref{sec:plres}), the mass
of $^{25}$Ne should be taken from {\sc AME'95} and not from {\sc
Ame2003} which includes this doubtful result. We draw the reader's
attention on the fact that {\sc Ame2003} also includes the
preliminary values for $^{26}$Ne and $^{29-33}$Mg, and consequently,
these values must be replaced by the present ones that are slightly
more precise and much more accurate (Table \ref{tab:summ}).

The two {\sc Mistral} mass measurements of $^{26}$Na ({\sc Thermo}
and {\sc Rilis} experiments) are in good agreement and in agreement
also with three other measurements using different methods. However
they are found to be in conflict with the result of Los Alamos
\cite{74Fl01}. It would therefore be interesting to remeasure this
mass with a Penning trap.

Finally, the three repeated measurements ($^{26}$Na in the {\sc
Thermo} and {\sc Rilis} experiments; $^{29}$Mg and $^{32}$Mg in the
{\sc Plasma} and {\sc Rilis} experiments) being in very good
agreement, show that {\sc Mistral} results suffer no extra
instrumental errors; and the overall good agrement of {\sc
Mistral}'s results with the existing ones, points out the accuracy
of {\sc Mistral} measurements.

   \section{Acknowledgements}

  We thank W.~Mittig for fruitful
discussions and H.~Doubre for assistance during the experiments.
   We would like to acknowledge the expert technical assistance of
G.~Conreur, M.~Jacotin, J.-F.~K\'epinski and G.~Le~Scornet from the
{\sc Csnsm}.
   {\sc Mistral} is supported by France's {\sc IN2P3}.
   The work at {\sc Isolde} was partially supported by the EU RTD
program ''Access to Research Infrastructures,'' under contract number
HPRI-CT-1998-00018 and the research network {\sc Nipnet} (contract
number HPRI-CT-2001-50034).

\clearpage

\clearpage
   \begin{table}
   \begin{center}
   \caption{\small \label{tab:CalMg}
   Calibration data from the {\sc Rilis} experiment
expressed as relative mass differences. The second part gives the
averaged results. Column~2 gives the measured value
$\Delta^{meas}_{x}$ and column~3, the value corrected for
calibration, $\Delta^{corr}_{x}$. The errors between square brackets
include only the statistical and the `{static}' ($3 \times 10^{-7}$)
errors, while the ones between parentheses include also the
`{dynamic}' error ($4 \times 10^{-7}$).
   } \vskip 0.1in
   \small{
   \begin{tabular}{crrr|rrr}
 \hline
 \hline
 & Nuclide    & $\Delta^{meas}_{x} \times 10^7$ & $\Delta^{corr}_{x} \times 10^7$ &
   Nuclide    & $\Delta^{meas}_{x} \times 10^7$ & $\Delta^{corr}_{x} \times 10^7$ \\
\hline
 \multicolumn{3}{l}{Individual calibrant masses} & & & & \\
 & $^{26}$Mg  & $-28.2[1.2](4.2)$  &  $ 4.0(4.2)$  & $^{24}$Mg  & $-71.3[0.7](4.1)$  &  $-3.6(4.1)$ \\
 & $^{24}$Mg  & $-70.3[1.0](4.1)$  &  $-2.6(4.1)$  & $^{23}$Na  & $-80.3[0.9](4.1)$  &  $ 6.1(4.1)$ \\
 & $^{25}$Mg  & $-53.8[0.9](4.1)$  &  $-4.3(4.1)$  & $^{24}$Mg  & $-66.9[0.7](4.1)$  &  $ 0.8(4.1)$ \\
 & $^{23}$Na  & $-86.0[0.6](4.0)$  &  $ 0.4(4.0)$  & $^{23}$Na  & $-89.3[0.7](4.1)$  &  $-2.9(4.1)$ \\
 & $^{26}$Mg  & $-25.7[0.8](4.1)$  &  $ 6.5(4.1)$  & $^{26}$Mg  & $-35.6[0.8](4.1)$  &  $-3.4(4.1)$ \\
 & $^{25}$Mg  & $-48.6[0.8](4.1)$  &  $ 0.9(4.1)$  & $^{28}$Mg  & $ -1.1[1.6](4.3)$  &  $-1.9(4.3)$  \\
  \hline
   \multicolumn{4}{l}{Averaged masses} \\
 & $^{23}$Na  &                    &   \multicolumn{1}{r}{$ 1.2(2.4)$} \\
 & $^{24}$Mg  &                    &   \multicolumn{1}{r}{$-1.8(2.4)$} \\
 & $^{25}$Mg  &                    &   \multicolumn{1}{r}{$-1.7(2.9)$}\\
 & $^{26}$Mg  &                    &   \multicolumn{1}{r}{$ 2.3(2.4)$}\\
 & $^{28}$Mg  &                    &   \multicolumn{1}{r}{$-1.9(4.3)$}\\
 \hline\hline
   \end{tabular}}
   \end{center}
   \end{table}

\begin{table}
\begin{center}
   \caption{\small \label{tab:ResRilis}
   {\sc Mistral} measured masses for the {\sc Rilis} experiment.
In column~2, 3 and 4 are respectively the measured, corrected
relative  and absolute mass differences. Column~5 and 6 give the
deduced mass excess (ME) in $\mu$u and in keV. Column 7 gives the
mass excess from the {\sc Ame'95} mass table \cite{Ame95}. In the
`individual measurements' part, the errors, between square brackets,
include only the statistical and the `static' ($3 \times 10^{-7}$)
errors, while the ones between parentheses include also the
`dynamic' error ($4 \times 10^{-7}$). In the `averaged masses' part,
the errors between parentheses include the `calibration' error
(Eq.\,(\ref{eq:errcal})) added after averaging.
   } \vskip 0.1in
\scriptsize{
   \begin{tabular}{crrrrrr}
   \hline\hline
   Nuclide    & $\Delta^{meas}_{x} \times 10^7$  & $\Delta^{corr}_{x} \times 10^7$  & $\delta m_x$ ($\mu$u)&ME($\mu$u)    &ME(keV)&AME'95(keV)\\
   \hline
   \multicolumn{6}{l}{Individual measurements} \\
    $^{30}$Mg & $ 26.3[0.7](4.1)$ & $ -5.4(4.1)$   & & & & \\
    $^{31}$Mg & $ 48.4[1.1](4.2)$ & $  2.1(4.2)$   & & & & \\
    $^{32}$Mg & $ 20.3[0.3](4.0)$ & $-40.2(4.0)$   & & & & \\
    $^{33}$Mg & $ -9.3[0.9](4.1)$ & $-83.5(4.1)$   & & & & \\
    $^{26}$Na & $-15.8[0.6](4.0)$ & $ 16.2(4.0)$   & & & & \\
    $^{30}$Mg & $ 27.5[1.1](4.2)$ & $ -4.2(4.2)$   & & & & \\
    $^{31}$Mg & $ 62.9[2.0](4.5)$ & $ 16.5(4.5)$   & & & & \\
    $^{30}$Mg & $ 30.7[1.4](4.3)$ & $ -1.0(4.3)$   & & & & \\
    $^{29}$Mg & $ 36.2[1.3](4.2)$ & $ 19.6(4.2)$   & & & & \\
   \hline
   \multicolumn{6}{l}{Averaged masses} \\
    $^{26}$Na &                   & $ 16.2(4.3)$   & $  42(11)$ & $ -7368(11)$ & $ -6863(11)$ & $- 6902(14) $\\
    $^{29}$Mg &                   & $ 19.6(5.5)$   & $  57(16)$ & $-11388(16)$ & $-10608(15)$ & $-10661(29) $\\
    $^{30}$Mg &                   & $ -3.6(4.8)$   & $ -11(14)$ & $ -9546(14)$ & $ -8892(13)$ & $ -8880(70) $\\
    $^{31}$Mg &                   & $  8.8(5.7)$   & $  27(18)$ & $ -3425(18)$ & $ -3190(16)$ & $ -3220(80) $\\
    $^{32}$Mg &                   & $-40.2(6.8)$   & $-129(22)$ & $  -983(22)$ & $  -915(20)$ & $  -800(100)$\\
    $^{33}$Mg &                   & $-83.5(7.3)$   & $-276(24)$ & $  5311(24)$ & $  4947(22)$ & $  5200(150)$\\

   \hline\hline
   \end{tabular}
   }
   \end{center}
  \end{table}

   \begin{table}
   \begin{center}
   \caption{\small \label{tab:CalMg99}
   Calibration data from the {\sc Plasma} experiment in chronological order,
expressed as relative mass differences.  The errors, between square
brackets, include only the statistical and the `{static}' ($5 \times
10^{-7}$) errors, while the ones between parentheses include also
the `dynamic' error ($7 \times 10^{-7}$).
   } \vskip 0.1in \footnotesize{
   \begin{tabular}{rrr|rrr}
 \hline
 \hline
    Nuclide-Ref.     &  time (min)  &  $\Delta^{meas}_{x} \times 10^7$
  & Nuclide-Ref.     &  time (min)  &  $\Delta^{meas}_{x} \times 10^7$
 \\\hline
 \multicolumn{2}{l}{Individual calibrant masses} & & &
 \\ $^{23}$Na-$^{23}$Na &  837 & $ 15.7 [ 0.5]( 7.0)$ & $^{26}$Mg-$^{23}$Na        & 3941 & $-42.3 [ 0.9]( 7.1)$
 \\ $^{23}$Ne-$^{23}$Na & 1166 & $ 17.6 [ 2.9]( 7.6)$ & $^{26}$Mg-$^{23}$Na        & 4233 & $-29.2 [ 1.2]( 7.1)$
 \\ $^{23}$Na-$^{23}$Na & 1216 & $ 17.7 [ 2.4]( 7.4)$ & $^{23}$Na-$^{23}$Na        & 4430 & $ 49.9 [ 0.8]( 7.0)$
 \\ $^{23}$Ne-$^{23}$Na & 1265 & $  4.5 [ 1.5]( 7.2)$ & $^{14}$C$^{15}$O-$^{23}$Na & 4896 & $-95.9 [ 4.2]( 8.1)$
 \\ $^{23}$Na-$^{23}$Na & 1307 & $ 18.3 [ 1.9]( 7.3)$ & $^{14}$N$^{15}$N-$^{23}$Na & 5045 & $-91.5 [ 4.3]( 8.2)$
 \\ $^{25}$Mg-$^{23}$Na & 1382 & $-21.8 [ 1.1]( 7.1)$ & $^{29}$Si-$^{23}$Na        & 5050 & $-90.1 [13.1](14.8)$
 \\ $^{25}$Mg-$^{23}$Na & 1404 & $-26.9 [ 1.0]( 7.1)$ & $^{23}$Na-$^{23}$Na        & 5180 & $ 44.7 [ 1.7]( 7.2)$
 \\ $^{25}$Na-$^{23}$Na & 1614 & $-35.2 [ 2.7]( 7.5)$ & $^{23}$Na-$^{23}$Na        & 5355 & $ 53.4 [ 0.9]( 7.1)$
 \\ $^{25}$Mg-$^{23}$Na & 1644 & $-31.1 [ 2.2]( 7.3)$ & $^{26}$Mg-$^{23}$Na        & 5430 & $-36.5 [ 2.0]( 7.3)$
 \\ $^{25}$Mg-$^{23}$Na & 1890 & $-28.9 [ 1.4]( 7.1)$ & $^{26}$Mg-$^{23}$Na        & 5520 & $-44.7 [ 1.9]( 7.3)$
 \\ $^{27}$Mg-$^{23}$Na & 2350 & $-57.7 [ 1.6]( 7.2)$ & $^{12}$C$^{18}$O-$^{41}$K  & 5776 & $159.0 [ 6.0]( 9.2)$
 \\ $^{27}$Mg-$^{23}$Na & 2384 & $-46.6 [ 2.5]( 7.4)$ & $^{12}$C$^{18}$O-$^{41}$K  & 5797 & $170.3 [ 5.8]( 9.1)$
 \\ $^{27}$Al-$^{23}$Na & 2499 & $-54.8 [ 1.5]( 7.2)$ & $^{15}$N$^{15}$N-$^{41}$K  & 5810 & $169.1 [ 4.0]( 8.0)$
 \\ $^{27}$Al-$^{23}$Na & 2548 & $-42.7 [ 1.3]( 7.1)$ & $^{30}$Ni-$^{41}$K         & 5978 & $153.6 [ 1.8]( 7.2)$
 \\ $^{27}$Al-$^{23}$Na & 2632 & $-39.6 [ 2.0]( 7.3)$ & $^{30}$Ni-$^{39}$K         & 5999 & $140.3 [ 2.3]( 7.4)$
 \\ $^{27}$Mg-$^{23}$Na & 2666 & $-46.1 [ 2.3]( 7.4)$ & $^{30}$Ni-$^{39}$K         & 6058 & $132.4 [ 4.6]( 8.4)$
 \\ $^{23}$Na-$^{23}$Na & 2918 & $ 20.4 [ 0.6]( 7.0)$ & $^{39}$Kr-$^{39}$K         & 6387 & $ 12.1 [ 4.9]( 8.6)$
 \\ $^{23}$Na-$^{23}$Na & 3011 & $ 29.4 [ 1.7]( 7.2)$ & $^{41}$Kr-$^{41}$K         & 6452 & $ 24.5 [ 3.8]( 8.0)$
 \\ $^{27}$Al-$^{23}$Na & 3039 & $-49.9 [ 1.3]( 7.1)$ & $^{39}$K -$^{41}$K         & 6990 & $ 22.0 [ 6.5](11.8)$
 \\ $^{27}$Mg-$^{23}$Na & 3052 & $-47.3 [ 2.1]( 7.3)$ & $^{30}$Ni-$^{41}$K         & 7170 & $157.8 [ 1.5]( 7.2)$
 \\ $^{27}$Mg-$^{23}$Na & 3400 & $-52.7 [ 1.4]( 7.1)$ & $^{30}$Ni-$^{39}$K         & 7247 & $135.8 [ 2.0]( 7.3)$
 \\ $^{23}$Na-$^{23}$Na & 3826 & $ 40.8 [ 2.7]( 7.5)$ & $^{32}$S -$^{39}$K         & 7280 & $ 91.5 [ 2.3]( 7.4)$
 \\                     &      &                      & $^{39}$K -$^{41}$K         & 7620 & $ 20.0 [12.1](15.6)$
 \\\hline\hline
   \end{tabular}
   }
   \end{center}
   \end{table}

   \begin{table}
   \begin{center}
   \caption{\small \label{tab:plasmacalib}
   Calibration results (for averaged masses) from the {\sc Plasma} experiment
expressed as relative mass differences.
   In column~3, are the measured value $\Delta_x{}_t^{meas}$ after taking into account the time dependance. The
   errors, between parentheses, include the statistical,
   the `{static}' ($5 \times 10^{-7}$), and the `{dynamic}' ($7 \times 10^{-7}$)    errors.
   Column~4 gives the value corrected for calibration,
$\Delta^{corr}_{x}$.
   Its error includes also the '{fitting}' error ($9 \times 10^{-7}$).
   } \vskip 0.1in
   \begin{tabular}{ccrr}
 \hline
 \hline
 {\sc Isolde} mass&{\sc Mistral} mass& $\Delta_x{}_t^{meas} \times 10^7$ & $\Delta^{corr}_{x} \times 10^7$ \\
\hline
 23        & 23        &   $14.5(2.2)$ &  $  8.1( 9.3)$ \\
 25        & 23        &  $-36.5(3.2)$ &  $ -2.0( 9.6)$ \\
 26        & 23        &  $-61.9(3.6)$ &  $ -8.2( 9.7)$ \\
 27        & 23        &  $-62.3(2.4)$ &  $  9.9( 9.3)$ \\
 29        & 23        & $-118.1(5.4)$ &  $-11.1(10.5)$ \\
 30        & 39        &  $104.1(4.4)$ &  $  0.4(10.0)$ \\
 30        & 41        &  $130.2(3.6)$ &  $  2.4( 9.7)$ \\
 39$^*$    & 41        &  $ 21.4(7.9)$ &  $ -3.2(12.0)$ \\
 $[41]^{**}$& $[41]^{**}$&  $ -13.4(5.8)$& $11.3(10.7)$ \\
 32        & 39        &   $55.1(7.4)$ &  $-17.3(11.6)$ \\
\hline\hline
\multicolumn{4}{l}{*  mass  from {\sc Mistral}}\\
\multicolumn{4}{l}{** mean value of $^{41}$K$^+$-$^{82}$Kr$^{++}$
and
$^{39}$K$^+$-$^{78}$Kr$^{++}$ measurements}\\
   \end{tabular}
   \end{center}
   \end{table}

   \begin{table}
   \begin{center}
   \caption{\small
   {\sc Mistral} measured masses for the {\sc Plasma} experiment using the new calibration law.
   In column~2 is the measured mass difference, in column 3 the
   averaged mass difference after taking into account the time
   dependance (Eqs. (\ref{eq:plcalib1}) and (\ref{eq:plcalib2})), in column
   4 the corrected relative mass difference (Eq. (\ref{eq:corr}))
   and in column 5 the absolute mass difference.
   In the `individual measurements' part,
   the error in column 2 includes only the statistical and the `static' errors ($5 \times 10^{-7}$),
   in column 3 the `dynamic' error ($7 \times 10^{-7}$) is added, and in column 4 the `fitting' error ($9 \times
   10^{-7}$) is added.
   In the `averaged masses' part,
   the error in columns 4 and 5 includes also the `calibration' error.
  } \vskip 0.1in
   \begin{tabular}{rcrrrrrrr}
   \hline\hline
 Nuclide-Ref.&  time(min) & $\Delta^{meas}_{x} \times 10^7$ &  $\Delta_x{}_t^{meas} \times 10^7$ &$\Delta^{corr}_{x} \times 10^7$ & $\delta m_x$ ($\mu$u)
  \\ \hline
   \multicolumn{5}{l}{Individual measurements}
  \\$^{29}$Mg-$^{23}$Na   & 4901 &$-73.2(3.2) $ & \multirow{2}{17.5mm}[.0mm]{ {\huge\}} $- 94(5)$} & \multirow{2}{10.5mm}[-.8mm]{$ 13(10)$}
  \\ $^{29}$Mg-$^{23}$Na  & 4943 &$-65.4(1.6)$
  \\ $^{26}$Ne-$^{23}$Na  & 5530 &$-18.9(1.6)$  & $ -47(7)$ &  $ 7(12)$
  \\ $^{32}$Mg-$^{39}$K   & 7540 &$ 78.0(10.3)$ & $  40(12)$& $-32(15)$
  \\ $^{32}$Mg-$^{41}$K   & 7700 &$ 98.0(7.8)$  & $  60(10)$& $-37(14)$
   \\\hline
   \multicolumn{5}{l}{Averaged masses}
\\$^{26}$Ne               &            &            &$  7(13)$ &  $18(34)$
\\$^{29}$Mg               &            &            &$ 13(13)$ &  $38(38)$
\\$^{32}$Mg               &            &            &$-35(12)$ &  $-112(38)$
\\\hline\hline
 \label{tab:mg99-res1}
  \end{tabular}
  \end{center}
  \end{table}

   \begin{table}
   \begin{center}
   \caption{\small
   {\sc Mistral} measured masses for the {\sc Plasma} experiment using the isobaric data.
   In columns~2 and 3 are the averaged measured relative mass differences for, respectively, the nuclide of interest
   and the average isobaric reference.
   Column 4 is the difference between the two preceding columns and is equivalent to $\Delta^{corr}_{x}$.
   Column 5 is the absolute mass difference.
   All the errors include the statistical, the `static' ($5 \times 10^{-7}$),
   and the `dynamic' ($6 \times 10^{-7}$) errors.
   } \vskip 0.1in
   \begin{tabular}{crrrrrr}
\hline\hline Nuclide-Ref.&  $\Delta^{meas}_{x} \times 10^7$
&$\Delta^{meas}_{r}\times 10^7$ &$\Delta_{x-r} \times 10^7$ &
$\delta m_x$ ($\mu$u)
\\\hline
$^{26}$Ne-26 &$-18.9(6.2)$ &$-40.6(4.5)$ & $22(8)$ & $ 56(20)$ \\
$^{29}$Mg-29        &$-69.0(4.6)$ &$-93.3(4.9)$ & $24(7)$ & $ 70(19)$ \\
   \hline\hline
\label{tab:mg99-res2}
   \end{tabular}
   \end{center}
   \end{table}

       \begin{table}
   \begin{center}
   \caption{\small
   {\sc Mistral} measured masses for the {\sc Plasma} experiment.
   In column 2 is the absolute mass difference.
   Columns~3 and 4 give the deduced mass excess (ME) in $\mu$u and in keV.
   The errors take into account all sources of errors.
   Column 5 gives the mass excess from the {\sc Ame'95} mass table \cite{Ame95}.
   } \vskip 0.1in
   \begin{tabular}{crrrrrr}
   \hline\hline
   Nuclide-Ref.& $\delta m_x$ ($\mu$u) &ME($\mu$u) & ME(keV)&AME'95(keV)
\\\hline
  $^{26}$Ne & $ 56(20)$ & $ 518(20)$   & $482(18)$    & $ 430(50) $
\\$^{29}$Mg & $ 70(19)$ & $-11375(19)$ & $-10596(18)$ & $-10661(29) $
\\$^{32}$Mg & $-112(38)$& $-966(38)$   & $-900(36)$   & $  -800(100)$
  \\ \hline\hline
\label{tab:mg99-res3}
   \end{tabular}
   \end{center}
   \end{table}

   \begin{table}
   \begin{center}
   \caption{\small
   Calibration parameters from the {\sc Thermo} experiment determined for each of the measurement periods,
after addition of the `fitting' error of 4$\times 10^{-7}$.
   In column 4, the correlation coefficients ($\sigma^2_{ab}$) between the fitted
parameters $a$ and $b$ are given.
   The $\chi$ values of the calibration fits are given in column 5. This table supersedes
   Table II of ref. \cite{Lun01a}.
   } \vskip 0.1in
   \begin{tabular}{l
   @{\hspace{.3cm}}r@{}c@{}l
   @{\hspace{.3cm}}r@{}c@{}l
   @{\hspace{.3cm}}r
   @{\hspace{.3cm}}r
   }
   \hline\hline
   Set
   &\multicolumn{3}{c}{Slope}
   &\multicolumn{3}{c}{Offset}
   &$\sigma^2_{ab}$
   &$\chi$\\
   &\multicolumn{3}{c}{($a \times 10^7$)}
   &\multicolumn{3}{c}{($b \times 10^7$)}\\
   \hline
   \#1a &$210. $&$\,($& 16.) &       &     &      &          & \\
   \#1b &$232. $&$\,($& 10.) &$ 1.7$ &$\,($&4.0)  &$ -33.5$  &0.5 \\
   \#2a &$ 26.5$&$\,($& 9.5) &$-4.6$ &$\,($&2.9)  &$ -14.3$  &0.4 \\
   \#2b &$ 83.1$&$\,($& 5.8) &$-4.5$ &$\,($&1.9)  &$  -6.6$  &1.1 \\
   \#2c &$-17.1$&$\,($& 4.6) &$-2.8$ &$\,($&1.7)  &$  -5.1$  &0.8 \\
   \#2d &$-35.0$&$\,($& 38.) &$-3.0$ &$\,($&3.9)  &$ 107. $  &0.2 \\
   \hline\hline
   \end{tabular}
   \label{tab:newfitNa}
   \end{center}
   \end{table}

  \begin{table}
  \begin{center}
   \caption{\small \label{tab:CalNa}
   Calibration results from the {\sc Thermo} experiment
expressed as relative mass differences. The second part gives the
averaged results. In column~2, the measured value
$\Delta^{meas}_{x}$ is shown. The error, between parentheses, is the
statistical error. Column~3 gives the value corrected for
calibration, $\Delta^{corr}_{x}$. Its error includes the `fitting'
error ($4 \times 10^{-7}$). This table supersedes Tables I+III+V of
ref. \cite{Lun01a}.}
   \vskip 0.1in \small{
   \begin{tabular}{lrr|lrr}
   \hline\hline
    Nuclide-Ref.   & $\Delta^{meas}_{x} \times 10^7$  & $\Delta^{corr}_{x} \times 10^7$
  & Nuclide-Ref.   & $\Delta^{meas}_{x} \times 10^7$  & $\Delta^{corr}_{x} \times 10^7$ \\
   \hline
   \multicolumn{2}{l}{Individual calibrant masses} & & & &\\
   \multicolumn{2}{l}{set\#1a} & & \multicolumn{3}{l}{set\#2c}\\
  $\quad ^{27}$Al-$^{39}$K  & $ 78.0(0.5)$& $  0.(4.0)$ & $\quad ^{23}$Na-$^{23}$Na & $ -1.5(1.3)$& $ 1.3(4.2)$ \\
   \multicolumn{2}{l}{set\#1b}  &                       & $\quad ^{23}$Na-$^{39}$K  & $ -7.1(1.4)$& $ 4.5(4.2)$\\
  $\quad ^{25}$Na-$^{39}$K  & $103.5(0.6)$& $ 0.4(4.0)$ & $\quad ^{25}$Na-$^{23}$Na & $ -2.8(1.5)$& $-1.4(4.3)$ \\
  $\quad ^{24}$Na-$^{39}$K  & $113.7(0.6)$& $ 1.6(4.0)$ & $\quad ^{25}$Na-$^{39}$K  & $ -8.3(1.9)$& $ 2.0(4.4)$ \\
  $\quad ^{23}$Na-$^{23}$Na & $  1.5(0.5)$& $-0.2(4.0)$ & $\quad ^{23}$Na-$^{23}$Na & $ -2.6(1.5)$& $ 0.2(4.3)$\\
  $\quad ^{23}$Na-$^{39}$K  & $119.5(0.7)$& $-1.8(4.1)$ & $\quad ^{23}$Na-$^{39}$K  & $ -8.0(1.4)$& $ 3.6(4.2)$ \\
   \multicolumn{2}{l}{set\#2a}  &                       & $\quad ^{24}$Na-$^{23}$Na & $ -4.1(1.3)$& $-2.0(4.2)$\\
  $\quad ^{23}$Na-$^{23}$Na & $ -3.9(2.4)$& $ 0.7(4.7)$ & $\quad ^{24}$Na-$^{39}$K  & $-15.2(1.4)$& $-4.3(4.2)$\\
  $\quad ^{25}$Na-$^{23}$Na & $ -7.1(1.3)$& $-0.3(4.2)$ & $\quad ^{27}$Al-$^{23}$Na & $  2.1(1.0)$& $ 2.2(4.1)$\\
  $\quad ^{25}$Na-$^{39}$K  & $  4.9(3.0)$& $-2.1(5.0)$ & $\quad ^{27}$Al-$^{39}$K  & $-10.9(1.1)$& $-1.9(4.2)$\\
   \multicolumn{2}{l}{set\#2b} & &    \multicolumn{3}{l}{set\#2d}\\
  $\quad ^{23}$Na-$^{23}$Na & $ -5.6(1.4)$& $-1.1(4.2)$ & $\quad ^{23}$Na-$^{23}$Na & $ -3.2(0.6)$& $-0.1(4.0)$\\
  $\quad ^{23}$Na-$^{39}$K  & $ 38.1(2.1)$& $-0.3(4.5)$ & $\quad ^{26}$Na-$^{23}$Na & $ 18.9(0.8)$& $17.7(4.4)$ \\
  $\quad ^{25}$Na-$^{23}$Na & $-13.3(1.9)$& $-1.9(4.4)$ & $\quad ^{28}$Na-$^{23}$Na & $ 20.4(6.0)$& $16.7(8.2)$\\
  $\quad ^{25}$Na-$^{39}$K  & $ 36.8(1.0)$& $ 4.9(4.1)$ & & &\\
  $\quad ^{27}$Al-$^{23}$Na & $-14.5(1.8)$& $ 3.3(4.4)$ & & &\\
  $\quad ^{27}$Al-$^{39}$K  & $ 17.8(1.1)$& $-7.9(4.2)$ & & &\\
  $\quad ^{23}$Na-$^{23}$Na & $ -3.6(1.6)$& $ 0.9(4.3)$ & & &\\
     \hline
   \multicolumn{3}{l}{Averaged masses} & & &\\
  $\quad ^{23}$Na           & & \multicolumn{1}{r}{$ 0.7(1.3)$}\\
  $\quad ^{24}$Na           & & \multicolumn{1}{r}{$-1.5(2.4)$}\\
  $\quad ^{25}$Na           & & \multicolumn{1}{r}{$ 0.4(1.6)$}\\
  $\quad ^{27}$Al           & & \multicolumn{1}{r}{$-1.1(2.1)$}\\
   \hline\hline
   \end{tabular}
   }
   \end{center}
  \end{table}

  \begin{table}
  \begin{center}
   \caption{\small \label{tab:ResNa1}
   {\sc Mistral} results from the {\sc Thermo} experiment
   expressed as relative mass differences.
   In columns~2 and 3 are, respectively, the measured and corrected relative mass difference.
  The second part, `** Averaged masses for set\#2c', gives the averaged results only for measurements which have
been repeated inside a set. The errors, in column 2, are the
statistical errors. In column~3, the errors, between square
brackets, include also the `fitting' error ($4 \times 10^{-7}$),
while for the ones between parentheses the `calibration' error is
added. This table supersedes Tables I+III of ref. \cite{Lun01a}.
   } \vskip 0.1in
\scriptsize{
   \begin{tabular}{lrr|lrr}
   \hline\hline
    Nuclide-Ref.   & $\Delta^{meas}_{x} \times 10^7$  & $\Delta^{corr}_{x} \times
    10^7$
  & Nuclide-Ref.   & $\Delta^{meas}_{x} \times 10^7$  & $\Delta^{corr}_{x} \times 10^7$  \\
   \hline
   \multicolumn{2}{l}{Individual masses} & & & &\\
   \multicolumn{2}{l}{set\#1a} & & \multicolumn{3}{l}{set\#2c}\\
  $\quad ^{27}$Na-$^{39}$K  & $103.0(1.2)$ &$ 25.0[ 4.2]( 7.1)$  &$\quad ^{28}$Na-$^{23}$Na &  $25.2(6.0)$ & $24.6[7.2](7.5)$ \\
  $\quad ^{28}$Na-$^{39}$K  & $ 91.2(4.5)$ &$ 20.6[ 6.0]( 8.0)$  &$\quad ^{28}$Na-$^{39}$K  &  $ 1.9(8.1)$ & $10.3[9.0](9.2)$\\
   \multicolumn{2}{l}{set\#1b}  &                                &$\quad ^{29}$Na-$^{23}$Na &  $26.1(6.2)$ & $25.0[7.4](7.7)$\\
  $\quad ^{29}$Na-$^{39}$K  & $ 95.5(7.0)$ &$ 25.6[ 8.1]( 8.9)$  &$\quad ^{29}$Na-$^{39}$K  &  $-0.1(6.3)$ & $ 7.7[7.5](7.7)$\\
  $\quad ^{26}$Na-$^{39}$K  & $100.0(0.5)$ &$  5.6[ 4.0]( 5.8)$  &$\quad ^{26}$Na-$^{23}$Na &  $20.1(1.0)$ & $20.8[4.1]**$\\
  $\quad ^{30}$Na-$^{39}$K  & $-13.2(9.7)$ &$-75.4[10.5](11.1)$  &$\quad ^{26}$Na-$^{39}$K  &  $12.1(1.0)$ & $21.7[4.1]**$\\
   \multicolumn{2}{l}{set\#2a}  &                                &$\quad ^{26}$Na-$^{23}$Na &  $21.3(1.5)$ & $22.0[4.3]**$\\
  $\quad ^{26}$Na-$^{23}$Na & $ 11.8(1.3)$ &$ 19.6[ 4.2]( 5.4)$  &$\quad ^{26}$Na-$^{39}$K  &  $10.3(1.5)$ & $19.9[4.3]**$\\
  $\quad ^{26}$Na-$^{39}$K  & $ 28.0(2.0)$ &$ 22.0[ 4.5]( 6.1)$  &$\quad ^{27}$Na-$^{23}$Na &  $23.1(2.2)$ & $23.2[4.6](5.0)$ \\
   \multicolumn{2}{l}{set\#2b} &                                 &$\quad ^{27}$Na-$^{39}$K  &  $17.2(1.4)$ & $26.2[4.2](4.7)$\\
  $\quad ^{26}$Na-$^{23}$Na & $ -7.7(1.9)$ &$  6.9[ 4.4]( 5.0)$  & \multicolumn{3}{l}{set\#2d}\\
  $\quad ^{26}$Na-$^{39}$K  & $ 43.7(2.6)$ &$ 15.0[ 4.8]( 5.4)$  &$\quad ^{30}$Na-$^{23}$Na & $-77.6(9.0)$ & $-83.7[9.9](13.7)$\\
  $\quad ^{28}$Na-$^{39}$K  & $ 33.0(9.1)$ &$ 10.2[ 9.9](10.2)$  && &\\
  \hline
     \multicolumn{3}{l}{** Averaged masses for set\#2c} \\
  $\quad ^{26}$Na-$^{23}$Na &              &\multicolumn{1}{r}{$ 21.4[3.0](3.5)$}\\
  $\quad ^{26}$Na-$^{39}$K  &              &\multicolumn{1}{r}{$ 20.9[3.0](3.6)$}\\
   \hline\hline
   \end{tabular}
   }
  \end{center}
  \end{table}

  \begin{table}
  \begin{center}
   \caption{\small \label{tab:ResNa2}
     {\sc Mistral} measured masses for the {\sc Thermo} experiment.
 In column~2 and 3 are respectively the corrected relative and
 absolute mass differences. Columns~4 and 5 give the deduced mass
 excess (ME) in $\mu$u and in keV. Column 6 gives the mass excess
 from the {\sc Ame'95} mass table \cite{Ame95}. The errors in columns
 2-5 take into account all sources of errors.
 This table supersedes Tables VI of ref. \cite{Lun01a}. }
 \vskip 0.1in
   \begin{tabular}{crrrrr}
   \hline\hline
Nuclide & $\Delta^{corr}_{x} \times 10^7$  & $\delta m_x$ ($\mu$u)& ME($\mu$u)    & ME(keV) & AME'95(keV)\\
   \hline
$^{26}$Na & $ 17.3(1.6)$ & $  45(4) $ & $-7365(4) $ & $-6860(4) $ & $-6902(14)$\\
$^{27}$Na & $ 25.2(1.5)$ & $  68(4) $ & $-5922(4) $ & $-5516(4) $ & $-5580(40)$\\
$^{28}$Na & $ 17.6(3.8)$ & $  49(11)$ & $-1061(11)$ & $ -988(10)$ & $-1030(80)$\\
$^{29}$Na & $ 18.9(4.6)$ & $  55(13)$ & $ 2866(13)$ & $ 2669(12)$ & $ 2620(90)$\\
$^{30}$Na & $-78.5(8.4)$ & $-236(25)$ & $ 8990(25)$ & $ 8375(23)$ & $ 8590(90)$\\
   \hline\hline
   \end{tabular}
  \end{center}
  \end{table}

\begin{table}
\begin{center}
   \caption{\small \label{tab:summ}
   Summary of the new reevaluated {\sc Mistral} results.
   } \vskip 0.1in
   \begin{tabular}{crrl}
   \hline\hline
   Nuclide    & ME($\mu$u)    & ME(keV) & Experiment\\
   \hline
    $^{26}$Na & $ -7368(11)$ & $ -6863(11)$ & {\sc Rilis} \\
    $^{29}$Mg & $-11388(16)$ & $-10608(15)$ & \\
    $^{30}$Mg & $ -9546(14)$ & $ -8892(13)$ & \\
    $^{31}$Mg & $ -3425(18)$ & $ -3190(16)$ &\\
    $^{32}$Mg & $  -983(22)$ & $  -915(20)$ & \\
    $^{33}$Mg & $  5311(24)$ & $  4947(22)$ & \\
    $^{26}$Ne & $   518(20)$ & $   482(18)$ & {\sc Plasma} \\
    $^{29}$Mg & $-11375(19)$ & $-10596(18)$ & \\
    $^{32}$Mg & $  -966(38)$ & $  -900(36)$ & \\
    $^{26}$Na & $ -7365( 4)$ & $ -6860( 4)$ & {\sc Thermo} \\
    $^{27}$Na & $ -5922( 4)$ & $ -5516( 4)$ &  \\
    $^{28}$Na & $ -1061(11)$ & $  -988(10)$ &  \\
    $^{29}$Na & $  2866(13)$ & $  2669(12)$ &  \\
    $^{30}$Na & $  8990(25)$ & $  8375(23)$ & \\
    \hline\hline
   \end{tabular}
   \end{center}
  \end{table}

    \begin{table}
  \begin{center}
   \caption{\small \label{tab:disc}
New adjustment of the data related to $^{26}$Na taking into account
our data from Table~\ref{tab:summ}. Column 2 is the measured value
in $\mu$u for mass-spectrometry and in keV for reactions and decays.
Column 3 is the output of a least-squares fit of all available
experimental data as in {\sc Ame'2003} (Ref. \cite{Ame03}, p. 184).
Column 4 gives the difference between the measured and adjusted
values relative to the uncertainty of the measured value.}
   \vskip 0.1in \footnotesize{
   \begin{tabular}{lrrrl}
   \hline\hline
  Measurement                                           & Measured value & Adjusted value & $v_i$& Reference\\
   \hline
  $^{26}$Na                                             &  $ -7365(4)  $ & $ -7365(4) $ & $-0.1$ &  {\sc Thermo} \\
  $^{26}$Na                                             &  $ -7368(11) $ & $ -7365(4) $ & $ 0.2$ &  {\sc Rilis}  \\
  $^{26}$Na($\beta^-$)$^{26}$Mg                         &  $  9210(200)$ & $  9354(4) $ & $ 0.7$ &  \cite{73Al13} \\
  $^{26}$Mg($^7$Li,$^7$Be)$^{26}$Na*                    &  $-10182(40) $ & $-10216(4) $ & $-0.8$ &  \cite{72Ba35} \\
  $^{26}$Mg($t$,$^3$He)$^{26}$Na                        &  $ -9292(20) $ & $ -9335(4) $ & $-2.2$ &  \cite{74Fl01} \\
  $^{25}$Na$-.721 \times ^{26}$Na$-.284 \times ^{22}$Na &  $ -2881(33) $ & $ -2940(4) $ & $-1.8$ &  \cite{75Th08} \\
  $^{25}$Na$-.721 \times ^{26}$Na$-.284 \times ^{22}$Na &  $ -2921(22) $ & $ -2940(4) $ & $-0.8$ &  \cite{75Th08} \\
  $^{26}$Na$-.770 \times ^{27}$Na$-.236 \times ^{22}$Na &  $ -1437(86) $ & $ -1389(5) $ & $ 0.6$ &  \cite{75Th08} \\
  $^{26}$Na$-.481 \times ^{27}$Na$-.520 \times ^{25}$Na &  $   676(66) $ & $   658(6) $ & $-0.2$ &  \cite{75Th08} \\
  $^{26}$Na$-.481 \times ^{27}$Na$-.520 \times ^{25}$Na &  $   734(86) $ & $   658(6) $ & $-0.6$ &  \cite{75Th08} \\
  $^{26}$Na$-.619 \times ^{28}$Na$-.394 \times ^{22}$Na &  $ -4229(613)$ & $ -4208(7) $ & $ 0.0$ &  \cite{75Th08} \\
  $^{26}$Na$-.619 \times ^{28}$Na$-.394 \times ^{22}$Na &  $ -4205(128)$ & $ -4208(7) $ & $ 0.0$ &  \cite{75Th08} \\
  $^{26}$Na$-.619 \times ^{28}$Na$-.394 \times ^{22}$Na &  $ -4203(87) $ & $ -4208(7) $ & $-0.1$ &  \cite{75Th08} \\
  $^{26}$Na$-.512 \times ^{29}$Na$-.506 \times ^{22}$Na &  $ -5763(91) $ & $ -5606(7) $ & $ 1.2$ &  \cite{75Th08} \\
  $^{26}$Na$-.512 \times ^{29}$Na$-.506 \times ^{22}$Na &  $ -6379(293)$ & $ -5606(7) $ & $ 1.8$ &  \cite{75Th08} \\
  $^{26}$Na$-.512 \times ^{29}$Na$-.506 \times ^{22}$Na &  $ -5252(277)$ & $ -5606(7) $ & $-0.5$ &  \cite{75Th08} \\
  $^{26}$Na$-.512 \times ^{29}$Na$-.506 \times ^{22}$Na &  $ -5576(66) $ & $ -5606(7) $ & $-0.5$ &  \cite{75Th08} \\
  $^{26}$Na$-.433 \times ^{30}$Na$-.591 \times ^{22}$Na &  $ -7454(287)$ & $ -7425(24)$ & $ 0.1$ &  \cite{75Th08} \\
  $^{26}$Na$-.433 \times ^{30}$Na$-.591 \times ^{22}$Na &  $ -8060(641)$ & $ -7425(24)$ & $ 0.7$ &  \cite{75Th08} \\
  $^{26}$Na$-.433 \times ^{30}$Na$-.591 \times ^{22}$Na &  $ -7045(225)$ & $ -7425(24)$ & $-0.7$ &  \cite{75Th08} \\
  $^{26}$Na$-.433 \times ^{30}$Na$-.591 \times ^{22}$Na &  $ -7515(117)$ & $ -7425(24)$ & $ 0.8$ &  \cite{75Th08} \\
   \hline\hline
\multicolumn{5}{l}{* The original result, Q$=-10222(30)$\,keV, has
been corrected (see text).}\\
   \end{tabular}
   }
  \end{center}
  \end{table}

\end{document}